\documentclass[a4paper]{article}

\pdfoutput=1

\usepackage[english]{babel}

\usepackage[protrusion=true,expansion=true]{microtype}
\usepackage[T1]{fontenc}

\usepackage{lmodern}

\usepackage{mathtools}
\usepackage{amssymb}
\usepackage{algorithm2e}

\usepackage{bm} 
\usepackage{upgreek} 
\usepackage{bbold} 

\usepackage{textcomp} 
\usepackage{gensymb} 

\usepackage{graphicx} 
\usepackage{float} 
\usepackage{placeins}
\usepackage{hyperref} 
\usepackage[margin=2.5cm]{geometry} 
\usepackage[labelfont=bf]{caption} 
\usepackage{subcaption}
\usepackage{authblk}                 

\usepackage{color} 
\usepackage{physics} 
\usepackage{rotating} 
\usepackage{longtable} 
\usepackage{cleveref} 

\usepackage{csquotes} 

\usepackage{makecell}
\usepackage{booktabs}
\usepackage{multirow}

\usepackage{diagbox}

\usepackage[style=numeric-comp,url=true,sorting=none]{biblatex}
\renewbibmacro*{doi+eprint+url}{%
    \printfield{doi}%
    \newunit\newblock%
    \iftoggle{bbx:eprint}{%
        \usebibmacro{eprint}%
    }{}%
    \newunit\newblock%
    \iffieldundef{doi}{%
        \usebibmacro{url+urldate}}%
        {}%
    }
\title{Self-consistent clustering analysis for homogenisation of heterogeneous plates}
\author[1]{Menglei Li\textsuperscript{*}}
\author[2]{Haolin Li\textsuperscript{*}}
\author[1]{Bing Wang\textsuperscript{†}}
\author[1]{Bing Wang\textsuperscript{‡}}
\affil[1]{National Key Laboratory of Science and Technology on Advanced Composites in Special Environments, Harbin Institute of Technology, Harbin 150001, PR China}
\affil[2]{Department of Aeronautics, Imperial College London, London, SW7 2AZ, UK}
\date{}

\begin{document}
\maketitle
\begingroup
\renewcommand\thefootnote{\fnsymbol{footnote}}
\footnotetext[1]{Menglei Li and Haolin Li contributed equally to this work.}
\footnotetext[2]{Third author: \href{mailto:wangbing91@hit.edu.cn}{wangbing91@hit.edu.cn}}
\footnotetext[3]{Corresponding author, \href{mailto:wangbing86@hit.edu.cn}{wangbing86@hit.edu.cn}}
\endgroup

{\abstract{\noindent}}
This work introduces a reduced-order model for plate structures with periodic micro-structures by coupling self-consistent clustering analysis (SCA) with the Lippmann–Schwinger equation, enabling rapid multiscale homogenisation of heterogeneous plates. A plate-specific SCA scheme is derived for the first time and features two key elements: (i) an offline–online strategy that combines Green’s functions with \emph{k}-means data compression, and (ii) an online self-consistent update that exploits the weak sensitivity of the reference medium. The framework handles both linear and non-linear problems in classical plate theory and first-order shear deformation theory, and its performance is verified on linear isotropic perforated plates and woven composites, as well as on non-linear elasto-plastic perforated plates and woven composites with damage. Across all cases the proposed model matches the accuracy of FFT-based direct numerical simulation while reducing computational cost by over an order of magnitude.


\section{Introduction}\label{sec:Introduction}

The growing demand for lightweight and high-performance materials in aerospace, automotive, and biomedical industries has led to major developments in multiscale heterogeneous materials \cite{Ref1, Ref2}. Engineered materials such as woven composites, porous metamaterials, and thin-walled plates with periodic microstructures achieve high stiffness-to-weight ratios and multiple functions through carefully designed microscopic periodic structures \cite{Ref3, Ref4}. However, their complex multiscale structures make mechanical characterisation difficult. Classical homogenisation theory, which builds equivalent mechanical models for representative volume elements (RVEs) at the microscale, has become a key approach for connecting the micro- and macro-scale mechanical behaviours in heterogeneous materials \cite{Ref5, Ref6}.

With progress in computational mechanics, numerical solutions to RVE boundary value problems have become the main approach. The finite element method (FEM), as the most established and commonly used numerical tool, remains the standard for homogenisation analysis \cite{Ref7, Ref8}. However, FEM has limitations when dealing with complex geometries and multiscale problems, as it faces rapidly increasing meshing difficulty and computational cost. The fast Fourier transform (FFT)-based homogenisation method, first proposed by Moulinec and Suquet \cite{Ref9}, brought a major change by reformulating the RVE problem into periodic Lippmann-Schwinger equations using the convolution theorem and Green’s operator. This method allows efficient solutions for heterogeneous material unit cells with subgrid-level accuracy, reduces computational cost to $O(N\log N)$, and avoids the need for complex meshing. Over the years, FFT-based methods have been applied successfully to many areas, including composite damage modelling \cite{Ref10}, crystal plasticity \cite{Ref11}, and multiphysics coupling analysis \cite{Ref12}. A full review of these developments and improvements can be found in \cite{Ref13, Ref14}.

Despite significant progress, FFT-based direct numerical simulation (DNS) remains constrained by voxel discretisation scale, posing challenges for ultra-large-scale and high-fidelity multiscale computations. To alleviate this issue, reduced-order models (ROMs) such as TFA \cite{Ref15}, NTFA \cite{Ref16}, POD \cite{Ref17}, PGD \cite{Ref18}, and ECM \cite{Ref19} have been developed to enhance computational efficiency. However, these methods typically rely on empirical assumptions or require extensive prior datasets to ensure predictive accuracy. The self-consistent clustering analysis (SCA) framework proposed by Liu et al. \cite{Ref20} represents a paradigm shift in data-driven reduced-order modelling. By incorporating mechanical response clustering into the FFT framework, SCA constructs cluster-based Lippmann-Schwinger equations that retain the computational accuracy of full-field simulations while achieving the efficiency of traditional ROMs. This cluster-based reduced-order model (CROM) has demonstrated remarkable success in critical applications, including progressive failure analysis \cite{Ref21}, multiphysics coupling \cite{Ref22}, and multiscale computation \cite{Ref23, Ref24, Ref25, Ref26}. Further advancements, such as virtual cluster analysis (VCA) \cite{Ref27} and finite cluster analysis (FCA) \cite{Ref28}, supported by rigorous mathematical convergence proofs \cite{Ref29}, underscore the transformative potential of CROM in the field of computational homogenisation \cite{Ref30, Ref31, Ref32}.

It is worth noting that existing homogenisation techniques predominantly address cell problems formulated within the framework of solid mechanics. However, lightweight engineering structures, such as aircraft skins \cite{Ref33} and automotive body panels \cite{Ref34}, exhibit distinct thin-walled plate or shell characteristics that necessitate homogenisation descriptions based on plate theory \cite{Ref35}. Classical plate theory provides an efficient theoretical foundation for plate homogenisation by establishing the mapping between in-plane stress/moment resultants and macroscopic strain/curvature through ABD stiffness matrices. While FEM remains the dominant computational tool in this domain, recent efforts have explored FFT-based alternatives. Notably, Li et al. \cite{Ref36} introduced a Lippmann-Schwinger equation-based plate homogenisation model, extending the FFT method to plate structures. However, this DNS-type approach remains computationally prohibitive for complex geometric configurations and multiscale analysis scenarios. Therefore, the development of CROM-based computational methodologies for plate models represents a crucial breakthrough, offering substantial theoretical and engineering benefits in multiscale computation \cite{Ref37}, topology optimisation \cite{Ref38}, and material database construction \cite{Ref39, Ref40, Ref41}.

In this study, we propose an innovative approach that systematically integrates the CROM framework into plate theory to establish a self-consistent clustering analysis methodology for thin-walled structures. A voxel-based finite prism discretisation scheme is employed to enable efficient numerical solutions for complex heterogeneous plate unit cells. The methodological scope is further expanded through the combined use of classical plate theory (CPT) and first-order shear deformation theory (FoPT). To address the coupling challenges between the reference medium and discrete wave vectors in FoPT, we introduce a physically justified decoupling hypothesis that removes redundant computations of interaction tensors, thereby ensuring the robustness and computational efficiency of SCA in moderately thick plate analyses.

The paper is structured as follows: Sec.~\ref{sec:Methodology} presents the proposed methodology, including the homogenisation equations based on plate theory, the construction of the cluster-based reduced-order model, and the numerical implementation. Sec.~\ref{sec:Results} verifies the accuracy of the method using benchmark cases covering linear elasticity, elasto-plasticity, and progressive failure. Sec.~\ref{sec:Discussion} analyses the convergence behaviour of reference medium parameters and evaluates the computational efficiency of the approach. Finally, Sec.~\ref{sec:conclusions} summarises the key findings and outlines possible directions for future research.

\section{Methodology}\label{sec:Methodology}

\subsection{Lippmann-Schwinger homogenization for heterogeneous plates}\label{sec:2.1}

\begin{figure}[htbp]
\centering
\includegraphics[width=16cm]{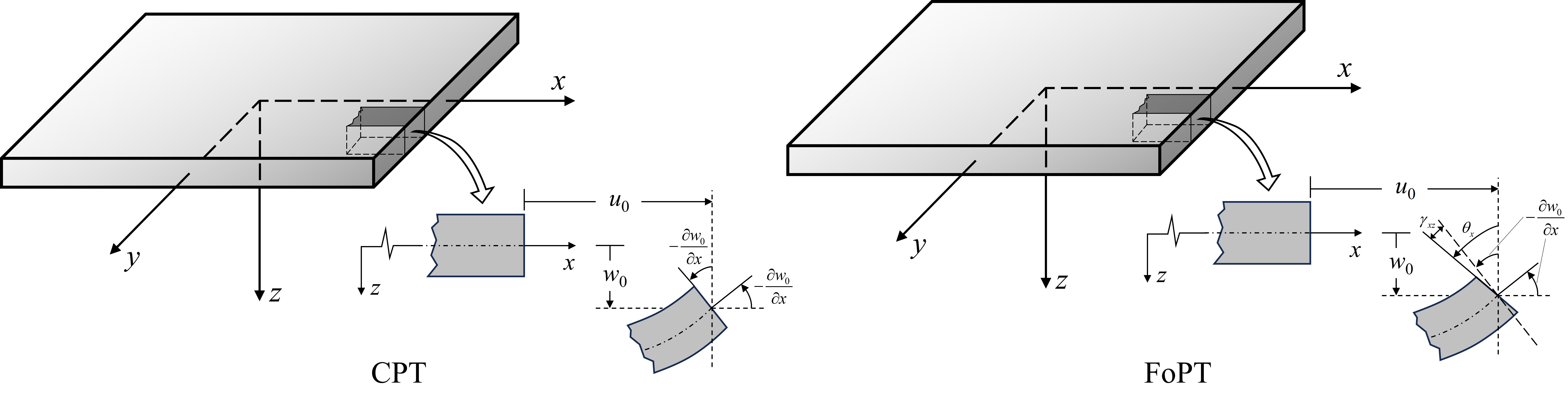}
\caption{\em{Definition of the degree of freedoms and the deformation assumption of the CPT and FoPT.}}
\label{fig:F1}
\end{figure}

In this study, we assume that the heterogeneous plate is thin and consider only  formulations of the Classical Plate Theory (CPT) and the First-Order Shear Deformation Theory (FoPT) \cite{Ref42}. The definitions of the degrees of freedom and the deformation assumptions for CPT and FoPT are shown in Fig.~\ref{fig:F1}. By defining $\Omega$ as the domain of the representative volume element (RVE) and $\boldsymbol{x}$ as a material point, the heterogeneous plate satisfies the following governing equation for the CPT problem:
\begin{equation} \label{eq:1}
\left\{
\begin{aligned}
  & \nabla \cdot \boldsymbol N(\boldsymbol x) = \boldsymbol 0, \quad \boldsymbol x \in \Omega  \\ 
  & \nabla \cdot \left[ \nabla \cdot \boldsymbol M(\boldsymbol x) \right] = \boldsymbol 0, \quad \boldsymbol x \in \Omega  \\ 
  & \boldsymbol N(\boldsymbol x) = \mathbb{A}(\boldsymbol x) : \boldsymbol \varepsilon(\boldsymbol x) + \mathbb{B}(x) : \boldsymbol \phi(\boldsymbol x), \quad \boldsymbol x \in \Omega  \\ 
  & \boldsymbol M(\boldsymbol x) = \mathbb{B}(\boldsymbol x) : \boldsymbol \varepsilon(\boldsymbol x) + \mathbb{D}(\boldsymbol x) : \boldsymbol \phi(\boldsymbol x), \quad \boldsymbol x \in \Omega  \\ 
  & \boldsymbol \varepsilon(\boldsymbol x) = \nabla \boldsymbol u(\boldsymbol x), \quad \boldsymbol x \in \Omega  \\ 
  & \boldsymbol \phi(\boldsymbol x) = \boldsymbol \Delta \boldsymbol w(\boldsymbol x), \quad \boldsymbol x \in \Omega  \\ 
\end{aligned}
\right.
\end{equation}
The heterogeneous plate satisfies the following governing equation for the FoPT problem:
\begin{equation} \label{eq:2}
\left\{ 
\begin{aligned}
  & \nabla \cdot \boldsymbol{N} \left( \boldsymbol{x} \right) = \boldsymbol 0, \quad \boldsymbol{x} \in \Omega  \\ 
  & \nabla \cdot \boldsymbol{M} \left( \boldsymbol{x} \right) - \boldsymbol{Q} \left( \boldsymbol{x} \right) = \boldsymbol 0, \quad \boldsymbol{x} \in \Omega  \\ 
  & \nabla \cdot \boldsymbol{Q} \left( \boldsymbol{x} \right) = \boldsymbol 0, \quad \boldsymbol{x} \in \Omega  \\ 
  & \boldsymbol{N} \left( \boldsymbol{x} \right) = \mathbb{A} \left( \boldsymbol{x} \right) : \boldsymbol{\varepsilon} \left( \boldsymbol{x} \right) + \mathbb{B} \left( \boldsymbol{x} \right) : \boldsymbol{\phi} \left( \boldsymbol{x} \right), \quad \boldsymbol{x} \in \Omega  \\ 
  & \boldsymbol{M} \left( \boldsymbol{x} \right) = \mathbb{B} \left( \boldsymbol{x} \right) : \boldsymbol{\varepsilon} \left( \boldsymbol{x} \right) + \mathbb{D} \left( \boldsymbol{x} \right) : \boldsymbol{\phi} \left( \boldsymbol{x} \right), \quad \boldsymbol{x} \in \Omega  \\ 
  & \boldsymbol{Q} \left( \boldsymbol{x} \right) = \boldsymbol{S} \left( \boldsymbol{x} \right) \cdot \boldsymbol{\gamma} \left( \boldsymbol{x} \right), \quad \boldsymbol{x} \in \Omega  \\ 
  & \boldsymbol{\varepsilon} \left( \boldsymbol{x} \right) = \nabla \boldsymbol{u} \left( \boldsymbol{x} \right), \quad \boldsymbol{x} \in \Omega  \\ 
  & \boldsymbol{\phi} \left( \boldsymbol{x} \right) = \nabla \boldsymbol{\theta} \left( \boldsymbol{x} \right), \quad \boldsymbol{x} \in \Omega  \\ 
  & \boldsymbol{\gamma} \left( \boldsymbol{x} \right) = \nabla \boldsymbol{w} \left( \boldsymbol{x} \right) + \boldsymbol{\theta} \left( \boldsymbol{x} \right), \quad \boldsymbol{x} \in \Omega  \\ 
\end{aligned}
\right.
\end{equation}
where $\boldsymbol{N}$,  $\boldsymbol{M}$, and  $\boldsymbol{Q}$ denote the stress, moment resultant and shear stress, respectively.  $\boldsymbol{\varepsilon}$, $\boldsymbol{\phi}$, and  $\boldsymbol{\gamma}$ represent the strain, curvature and shear strain, respectively. $\mathbb{A}$, $\mathbb{B}$,  $\mathbb{D}$, and $\boldsymbol{S}$ are the membrane stiffness, membrane-bending coupling stiffness, bending stiffness, and shear stiffness, respectively. $\boldsymbol{u}$, $\boldsymbol{\theta}$, and $\boldsymbol w$ denote the displacement, rotation angle and out-plane displacement, respectively.

The local displacement and rotation angle of the RVE model are decomposed into average and perturbation components:
\begin{equation} \label{eq:3}
\left\{ 
\begin{aligned}
  & {u}_{i} = {u}_{i}^{*} + {\bar{\varepsilon}}_{ij} {x}_{j}, \quad \text{and} \quad i \in 1,2,3, \quad j \in 1,2,3 \\ 
  & {\theta}_{i} = {\theta}_{i}^{*} + {\bar{\phi}}_{ij} {\theta}_{j}, \quad \text{and} \quad i \in 1,2, \quad j \in 1,2 \\ 
\end{aligned}
\right.
\end{equation}
Here, ${\bar{\varepsilon}}_{ij}$, ${\bar{\phi}}_{ij}$, ${u}_{i}^{*}$, and ${\theta}_{i}^{*}$ represent the average strain, average curvature, perturbation displacement, and perturbation rotation, respectively. The RVE model for heterogeneous plates satisfies periodic boundary conditions, which differ from those used in solid formulations, as they must also account for the periodicity of the rotational degrees of freedom. The periodic boundary conditions are expressed as follows:
\begin{equation} \label{eq:4}
\left\{ 
\begin{aligned}
  & u_{i} = u_{i}^{*} + \bar{\varepsilon}_{ij} x_{j}, \quad i \in 1,2, \quad j \in 1,2, \quad u_{i}^{*} \# \\ 
  & u_{3} = u_{3}^{*} + \bar{\gamma}_{3j} x_{j}, \quad j \in 1,2, \quad u_{3}^{*} \# \\ 
  & \theta_{i} = \theta_{i}^{*} + \bar{\phi}_{ij} \theta_{j}, \quad i \in 1,2, \quad j \in 1,2, \quad \theta_{i}^{*} \# \\ 
\end{aligned}
\right.
\end{equation}
where $\#$ denotes periodicity.

By introducing a homogeneous isotropic linear elastic reference medium (with stiffness $\mathbb{A}^0$, $\mathbb{B}^0$, $\mathbb{D}^0$, and $\boldsymbol{S}^0$), the boundary value problems (BVPs) of the RVE model for the heterogeneous plate can be solved using Green's function. First, the polarisation stress $\boldsymbol{\tau}^{(N)}$, polarisation moment $\boldsymbol{\tau}^{(M)}$, and polarisation shear stress $\boldsymbol{\tau}^{(Q)}$ are introduced as follows:
\begin{equation} \label{eq:5}
\left\{ 
\begin{aligned}
  & \boldsymbol{\tau}^{(N)}\left( \boldsymbol{x} \right) = \boldsymbol{N} \left( \boldsymbol{x} \right) - \left[ \mathbb{A}^{0} : \boldsymbol{\varepsilon} \left( \boldsymbol{x} \right) + \mathbb{B}^{0} : \boldsymbol{\phi} \left( \boldsymbol{x} \right) \right], \quad \boldsymbol{x} \in \Omega  \\ 
  & \boldsymbol{\tau}^{(M)}\left( \boldsymbol{x} \right) = \boldsymbol{M} \left( \boldsymbol{x} \right) - \left[ \mathbb{B}^{0} : \boldsymbol{\varepsilon} \left( \boldsymbol{x} \right) + \mathbb{D}^{0} : \boldsymbol{\phi} \left( \boldsymbol{x} \right) \right], \quad \boldsymbol{x} \in \Omega  \\ 
  & \boldsymbol{\tau}^{(Q)}\left( \boldsymbol{x} \right) = \boldsymbol{Q} \left( \boldsymbol{x} \right) - \boldsymbol{S}^{0} \cdot \boldsymbol{\gamma} \left( \boldsymbol{x} \right), \quad \boldsymbol{x} \in \Omega  \\ 
\end{aligned}
\right.
\end{equation}

The polarisation stress is treated as an external stress on the reference medium, and the original BVPs of CPT can be rewritten as the following Lippmann–Schwinger integral equations:
\begin{equation} \label{eq:6}
\left\{ 
\begin{aligned}
  & \boldsymbol{\varepsilon} \left( \boldsymbol{x} \right) + \int_{\Omega} \mathbb{G}^{0\left( NN \right)} \left( \boldsymbol{x} - \boldsymbol{y} \right) : \boldsymbol{\tau}^{\left( N \right)} \left( \boldsymbol{y} \right) d\boldsymbol{y} - \bar{\boldsymbol{\varepsilon}} = 0 \\ 
  & \boldsymbol{\phi} \left( \boldsymbol{x} \right) + \int_{\Omega} \mathbb{G}^{0\left( MM \right)} \left( \boldsymbol{x} - \boldsymbol{y} \right) : \boldsymbol{\tau}^{\left( M \right)} \left( \boldsymbol{y} \right) d\boldsymbol{y} - \bar{\boldsymbol{\phi}} = 0 \\ 
\end{aligned}
\right.
\end{equation}

The original BVPs of FoPT can be rewritten as the following Lippmann–Schwinger integral equations:
\begin{equation} \label{eq:7}
\left\{ 
\begin{aligned}
  & \boldsymbol{\varepsilon} \left( \boldsymbol{x} \right) + \int_{\Omega} \mathbb{G}^{0\left( NN \right)} \left( \boldsymbol{x} - \boldsymbol{y} \right) : \boldsymbol{\tau}^{\left( N \right)} \left( \boldsymbol{y} \right) d\boldsymbol{y} - \bar{\boldsymbol{\varepsilon}} = 0 \\ 
  & \boldsymbol{\phi} \left( \boldsymbol{x} \right) + \int_{\Omega} \mathbb{G}^{0\left( MM \right)} \left( \boldsymbol{x} - \boldsymbol{y} \right) : \boldsymbol{\tau}^{\left( M \right)} \left( \boldsymbol{y} \right) d\boldsymbol{y} + \int_{\Omega} \mathcal{G}^{0\left( MQ \right)} \left( \boldsymbol{x} - \boldsymbol{y} \right) : \boldsymbol{\tau}^{\left( Q \right)} \left( \boldsymbol{y} \right) d\boldsymbol{y} - \bar{\boldsymbol{\phi}} = 0 \\ 
  & \boldsymbol{\gamma} \left( \boldsymbol{x} \right) + \int_{\Omega} \mathcal{G}^{0\left( QM \right)} \left( \boldsymbol{x} - \boldsymbol{y} \right) : \boldsymbol{\tau}^{\left( M \right)} \left( \boldsymbol{y} \right) d\boldsymbol{y} + \int_{\Omega} \boldsymbol G^{0\left( QQ \right)} \left( \boldsymbol{x} - \boldsymbol{y} \right) : \boldsymbol{\tau}^{\left( Q \right)} \left( \boldsymbol{y} \right) d\boldsymbol{y} - \bar{\boldsymbol{\gamma}} = 0 \\ 
\end{aligned}
\right.
\end{equation}

$\mathbb{G}^{0\left( NN \right)}$, $\mathbb{G}^{0\left( MM \right)}$, $\mathcal{G}^{0\left( MQ \right)}$, $\mathcal{G}^{0\left( QM \right)}$, and $\boldsymbol G^{0\left( QQ \right)}$ represent different Green's operators. The Green's operators for CPT and FoPT have explicit formulations in Fourier space, as detailed in Appendix   \ref{sec:A.1} .

In addition, the Lippmann–Schwinger equations can be solved iteratively using the fixed-point iteration method \cite{Ref36}, which follows the basic framework of the Fourier spectral (FFT) method \cite{Ref9}. This approach is used as the DNS scheme in the present study.

\subsection{Solution framework based on clustered reduced-order modeling}\label{sec:2.2}

Although the solution of Eq.(\ref{eq:7}) using FFT can effectively improve computational speed, its efficiency still depends directly on the level of discretisation and thus remains within the domain of DNS. This approach continues to face computational challenges when applied to models with complex heterogeneity. Following the concept of cluster-based reduced-order models (CROMs), the RVE model can be decomposed into a finite number of material clusters. By deriving the cluster-based Lippmann–Schwinger equations, the computational degrees of freedom can be significantly reduced, resulting in improved efficiency.

Due to the complex coupling of Green's operators involved in FoPT, both the discretised Lippmann–Schwinger equation and the self-consistent update scheme differ significantly from the classical self-consistent clustering analysis (SCA) \cite{Ref20}. In contrast, the Lippmann–Schwinger equation in CPT involves only two simple Green's operators, and its derivation closely follows that of classical SCA. For convenience, the derivation for FoPT is presented below, while the derivation for CPT is provided in Appendix~\ref{sec:A.2}. First, the Lippmann–Schwinger equation, Eq.~(\ref{eq:7}), is written in an incremental form:
\begin{equation} \label{eq:8}
\left\{ 
\begin{aligned}
  & \Delta \boldsymbol{\varepsilon} \left( \boldsymbol{x} \right) + \int_{\Omega} \mathbb{G}^{0\left( NN \right)} \left( \boldsymbol{x} - \boldsymbol{y} \right) : \left[ \Delta \boldsymbol{N} \left( \boldsymbol{y} \right) - \mathbb{A}^{0} : \Delta \boldsymbol{\varepsilon} \left( \boldsymbol{y} \right) \right] d\boldsymbol{y} - \Delta \bar{\boldsymbol{\varepsilon}} = 0 \\ 
  & \Delta \boldsymbol{\phi} \left( \boldsymbol{x} \right) + \int_{\Omega} \mathbb{G}^{0\left( MM \right)} \left( \boldsymbol{x} - \boldsymbol{y} \right) : \left[ \Delta \boldsymbol{M} \left( \boldsymbol{y} \right) - \mathbb{D}^{0} : \Delta \boldsymbol{\phi} \left( \boldsymbol{y} \right) \right] d\boldsymbol{y} \\
  & \quad + \int_{\Omega} \mathcal{G}^{0\left( MQ \right)} \left( \boldsymbol{x} - \boldsymbol{y} \right) : \left[ \Delta \boldsymbol{Q} \left( \boldsymbol{y} \right) - \mathbb{S}^{0} \cdot \Delta \boldsymbol{\gamma} \left( \boldsymbol{y} \right) \right] d\boldsymbol{y} - \Delta \bar{\boldsymbol{\phi}} = 0 \\ 
  & \Delta \boldsymbol{\gamma} \left( \boldsymbol{x} \right) + \int_{\Omega} \mathcal{G}^{0\left( QM \right)} \left( \boldsymbol{x} - \boldsymbol{y} \right) : \left[ \Delta \boldsymbol{M} \left( \boldsymbol{y} \right) - \mathbb{D}^{0} : \Delta \boldsymbol{\phi} \left( \boldsymbol{y} \right) \right] d\boldsymbol{y} \\
  & \quad + \int_{\Omega} G^{0\left( QQ \right)} \left( \boldsymbol{x} - \boldsymbol{y} \right) : \left[ \Delta \boldsymbol{Q} \left( \boldsymbol{y} \right) - \mathbb{S}^{0} \cdot \Delta \boldsymbol{\gamma} \left( \boldsymbol{y} \right) \right] d\boldsymbol{y} - \Delta \bar{\boldsymbol{\gamma}} = 0 \\ 
\end{aligned}
\right.
\end{equation}

For the continuous form incremental equation, Eq.~(\ref{eq:8}), discretisation is required for numerical solution. It is assumed that the equation has been discretised into regular pixels, and the RVE model is divided into $H$ material clusters. The volume averaging of Eq.~(\ref{eq:8}) is performed within the \textit{I}-th cluster as follows:
\begin{equation} \label{eq:9}
\left\{ 
\begin{aligned}
  & \frac{1}{{c^{I}}|\Omega|} \int_{\Omega} \chi^{I} \left( \boldsymbol{x} \right) \Delta \boldsymbol{\varepsilon} \left( \boldsymbol{x} \right) d\boldsymbol{x} \\ 
  & + \frac{1}{{c^{I}}|\Omega|} \iint_{\Omega} \chi^{I} \left( \boldsymbol{x} \right) \mathbb{G}^{0\left( NN \right)} \left( \boldsymbol{x} - \boldsymbol{y} \right) : \left[ \Delta \boldsymbol{N} \left( \boldsymbol{y} \right) - \mathbb{A}^{0} : \Delta \boldsymbol{\varepsilon} \left( \boldsymbol{y} \right) \right] d\boldsymbol{y} d\boldsymbol{x} - \Delta \bar{\boldsymbol{\varepsilon}} = 0 \\ 
  & \frac{1}{{c^{I}}|\Omega|} \int_{\Omega} \chi^{I} \left( \boldsymbol{x} \right) \Delta \boldsymbol{\phi} \left( \boldsymbol{x} \right) d\boldsymbol{x} \\ 
  & + \frac{1}{{c^{I}}|\Omega|} \iint_{\Omega} \chi^{I} \left( \boldsymbol{x} \right) \mathbb{G}^{0\left( MM \right)} \left( \boldsymbol{x} - \boldsymbol{y} \right) : \left[ \Delta \boldsymbol{M} \left( \boldsymbol{y} \right) - \mathbb{D}^{0} : \Delta \boldsymbol{\phi} \left( \boldsymbol{y} \right) \right] d\boldsymbol{y} d\boldsymbol{x} \\ 
  & + \frac{1}{{c^{I}}|\Omega|} \iint_{\Omega} \chi^{I} \left( \boldsymbol{x} \right) \mathcal{G}^{0\left( MQ \right)} \left( \boldsymbol{x} - \boldsymbol{y} \right) : \left[ \Delta \boldsymbol{Q} \left( \boldsymbol{y} \right) - \mathbb{S}^{0} \cdot \Delta \boldsymbol{\gamma} \left( \boldsymbol{y} \right) \right] d\boldsymbol{y} d\boldsymbol{x} - \Delta \bar{\boldsymbol{\phi}} = 0 \\ 
  & \frac{1}{{c^{I}}|\Omega|} \int_{\Omega} \chi^{I} \left( \boldsymbol{x} \right) \Delta \boldsymbol{\gamma} \left( \boldsymbol{x} \right) d\boldsymbol{x} \\ 
  & + \frac{1}{{c^{I}}|\Omega|} \iint_{\Omega} \chi^{I} \left( \boldsymbol{x} \right) \mathcal{G}^{0\left( QM \right)} \left( \boldsymbol{x} - \boldsymbol{y} \right) : \left[ \Delta \boldsymbol{M} \left( \boldsymbol{y} \right) - \mathbb{D}^{0} : \Delta \boldsymbol{\phi} \left( \boldsymbol{y} \right) \right] d\boldsymbol{y} d\boldsymbol{x} \\ 
  & + \frac{1}{{c^{I}}|\Omega|} \iint_{\Omega} \chi^{I} \left( \boldsymbol{x} \right) G^{0\left( QQ \right)} \left( \boldsymbol{x} - \boldsymbol{y} \right) : \left[ \Delta \boldsymbol{Q} \left( \boldsymbol{y} \right) - \mathbb{S}^{0} \cdot \Delta \boldsymbol{\gamma} \left( \boldsymbol{y} \right) \right] d\boldsymbol{y} d\boldsymbol{x} - \Delta \bar{\boldsymbol{\gamma}} = 0 \\ 
\end{aligned}
\right.
\end{equation}
where $c^{I}$ denotes the volume fraction in the \textit{I}-th cluster. $ \chi^{I}$ is the characteristic function, defined as:
\begin{equation} \label{eq:10}
{\chi}^{I} \left( \boldsymbol{x} \right) = \left\{
\begin{aligned}
  & 1, \quad \boldsymbol{x} \in \Omega^{I} \\
  & 0, \quad \boldsymbol{x} \notin \Omega^{I}
\end{aligned}
\right.
\end{equation}

Assuming that the variable field is uniform in each region, the strain field of the RVE can be regarded as a piecewise linear interpolation of the strain fields within the individual material clusters:
\begin{equation} \label{eq:11}
\left\{ \begin{aligned}
  & \Delta \boldsymbol{\varepsilon} \left( \boldsymbol{x} \right) = \sum\limits_{I=1}^{H} {\chi}^{I} \left( \boldsymbol{x} \right) \Delta \boldsymbol{\varepsilon}^{I} \left( \boldsymbol{x} \right) \\
  & \Delta \boldsymbol{\phi} \left( \boldsymbol{x} \right) = \sum\limits_{I=1}^{H} {\chi}^{I} \left( \boldsymbol{x} \right) \Delta \boldsymbol{\phi}^{I} \left( \boldsymbol{x} \right) \\
  & \Delta \boldsymbol{\gamma} \left( \boldsymbol{x} \right) = \sum\limits_{I=1}^{H} {\chi}^{I} \left( \boldsymbol{x} \right) \Delta \boldsymbol{\gamma}^{I} \left( \boldsymbol{x} \right)
\end{aligned}
\right.
\end{equation}
Similarly, the stress field can be expressed as:
\begin{equation} \label{eq:12}
\left\{ \begin{aligned}
  & \Delta \boldsymbol{N} \left( \boldsymbol{x} \right) = \sum\limits_{I=1}^{H} {\chi}^{I} \left( \boldsymbol{x} \right) \Delta \boldsymbol{N}^{I} \left( \boldsymbol{x} \right) \\
  & \Delta \boldsymbol{M} \left( \boldsymbol{x} \right) = \sum\limits_{I=1}^{H} {\chi}^{I} \left( \boldsymbol{x} \right) \Delta \boldsymbol{M}^{I} \left( \boldsymbol{x} \right) \\
  & \Delta \boldsymbol{Q} \left( \boldsymbol{x} \right) = \sum\limits_{I=1}^{H} {\chi}^{I} \left( \boldsymbol{x} \right) \Delta \boldsymbol{Q}^{I} \left( \boldsymbol{x} \right)
\end{aligned}
\right.
\end{equation}
Substituting the strain and stress fields into Eq. ( \ref{eq:9} ):
\begin{equation} \label{eq:13}
\left\{ \begin{aligned}
  & \Delta \boldsymbol{\varepsilon}^{I} + \sum\limits_{J=1}^{H} \boldsymbol{D}^{IJ\left( NN \right)} \left[ \Delta \boldsymbol{N}^{J} - \mathbb{A}^{0} : \Delta \boldsymbol{\varepsilon}^{J} \right] - \Delta \bar{\boldsymbol{\varepsilon}} = 0 \\
  & \Delta \boldsymbol{\phi}^{I} + \sum\limits_{J=1}^{H} \boldsymbol{D}^{IJ\left( MM \right)} \left[ \Delta \boldsymbol{M}^{J} - \mathbb{D}^{0} : \Delta \boldsymbol{\phi}^{J} \right] + \sum\limits_{J=1}^{H} \boldsymbol{D}^{IJ\left( MQ \right)} \left[ \Delta \boldsymbol{Q}^{J} - \boldsymbol S^{0} \cdot \Delta \boldsymbol{\gamma}^{J} \right] - \Delta \bar{\boldsymbol{\phi}} = 0 \\
  & \Delta \boldsymbol{\gamma}^{I} + \sum\limits_{J=1}^{H} \boldsymbol{D}^{IJ\left( QM \right)} \left[ \Delta \boldsymbol{M}^{J} - \mathbb{D}^{0} : \Delta \boldsymbol{\phi}^{J} \right] + \sum\limits_{J=1}^{H} \boldsymbol{D}^{IJ\left( QQ \right)} \left[ \Delta \boldsymbol{Q}^{J} - \boldsymbol S^{0} \cdot \Delta \boldsymbol{\gamma}^{J} \right] - \Delta \bar{\boldsymbol{\gamma}} = 0
\end{aligned}
\right.
\end{equation}
Here, $\boldsymbol{D}^{IJ\left( \sim,\sim \right)}$ represents the interaction tensor, which characterizes the interaction between different clusters. The corresponding expression is as follows:
\begin{equation} \label{eq:14}
\left\{ \begin{aligned}
  & \boldsymbol{D}^{IJ\left( NN \right)} = \frac{1}{{{c}^{I}}\left| \Omega \right|} \iint_{\Omega} {\chi}^{I}(\boldsymbol{x}) {\chi}^{J}(\boldsymbol{x}) \mathbb{G}^{0\left( NN \right)}(\boldsymbol{x}-\boldsymbol{y}) \, dy \, dx \\
  & \boldsymbol{D}^{IJ\left( MM \right)} = \frac{1}{{{c}^{I}}\left| \Omega \right|} \iint_{\Omega} {\chi}^{I}(\boldsymbol{x}) {\chi}^{J}(\boldsymbol{x}) \mathbb{G}^{0\left( MM \right)}(\boldsymbol{x}-\boldsymbol{y}) \, dy \, dx \\
  & \boldsymbol{D}^{IJ\left( MQ \right)} = \frac{1}{{{c}^{I}}\left| \Omega \right|} \iint_{\Omega} {\chi}^{I}(\boldsymbol{x}) {\chi}^{J}(\boldsymbol{x}) \mathcal{G}^{0\left( MQ \right)}(\boldsymbol{x}-\boldsymbol{y}) \, dy \, dx \\
  & \boldsymbol{D}^{IJ\left( QM \right)} = \frac{1}{{{c}^{I}}\left| \Omega \right|} \iint_{\Omega} {\chi}^{I}(\boldsymbol{x}) {\chi}^{J}(\boldsymbol{x}) \mathcal{G}^{0\left( QM \right)}(\boldsymbol{x}-\boldsymbol{y}) \, dy \, dx \\
  & \boldsymbol{D}^{IJ\left( QQ \right)} = \frac{1}{{{c}^{I}}\left| \Omega \right|} \iint_{\Omega} {\chi}^{I}(x\boldsymbol{x}) {\chi}^{J}(\boldsymbol{x}) G^{0\left( QQ \right)}(\boldsymbol{x}-\boldsymbol{y}) \, dy \, dx \\
\end{aligned} \right.
\end{equation}

Therefore, in the cluster-based ROM, the focus is on solving the interaction tensor and the incremental Eq.~(\ref{eq:13}). The computation of Eq.~(\ref{eq:14}) relies on the material cluster information from the initial partition, corresponding to the "\textit{offline}" stage. The solution of Eq.~(\ref{eq:13}) depends on real-time loading and local material information, corresponding to the "\textit{online}" stage.

\subsection{Numerical implementation}\label{sec:2.3}

\subsubsection{Discretization} \label{sec:2.3.1}

\begin{figure}[htbp]
\centering
\includegraphics[width=16cm]{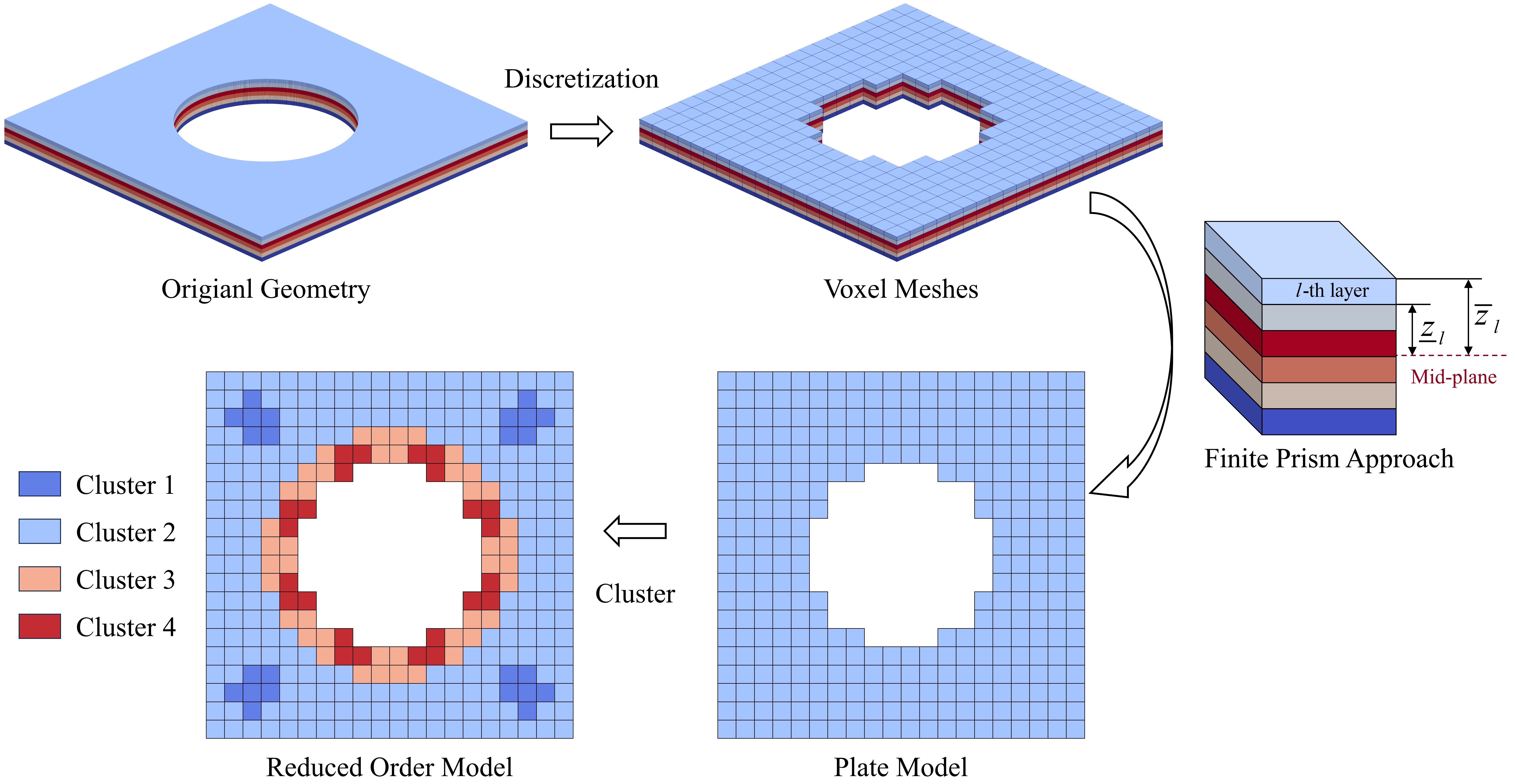}
\caption{\em{Schematic diagram of model discretization and order reduction.}}
\label{fig:F2}
\end{figure}

The simulation domain consists of a periodic RVE of the microstructure embedded in a cuboidal domain $\Omega$ with dimensions $L_1$, $L_2$, and $L_3$. In solid formulations, the RVE is discretised with a regular array of $N_1 \times N_2 \times N_3$ voxels, where each voxel belongs to one of the represented phases. When applying plate theory, the 3D solid structure is simplified into a 2D plate. In this study, a voxel-based finite prism approach is employed, where the discretised 3D architecture of the regular voxel array is represented as finite prisms \cite{Ref42}, as shown in Fig.~\ref{fig:F2}. In the thickness direction, the mid-plane of the plate is chosen as the origin of the coordinate system. The stiffnesses are calculated along the plate thickness based on classical laminate theory:
\begin{equation} \label{eq:15}
\left\{ \begin{aligned}
  & \mathbb{A} = \int_{z} \bar{\mathbb{Q}}_{m} \, dz = \sum_{l=1}^{n_{l}} \left[ \bar{\mathbb{Q}}_{m}^{(l)} \left( \bar{z}_{l} - \underline{z}_{l} \right) \right] \\
  & \mathbb{B} = \int_{z} z \bar{\mathbb{Q}}_{m} \, dz = \frac{1}{2} \sum_{l=1}^{n_{l}} \left[ \bar{\mathbb{Q}}_{m}^{(l)} \left( \bar{z}_{l}^{2} - \underline{z}_{l}^{2} \right) \right] \\
  & \mathbb{D} = \int_{z} z^{2} \bar{\mathbb{Q}}_{m} \, dz = \frac{1}{3} \sum_{l=1}^{n_{l}} \left[ \bar{\mathbb{Q}}_{m}^{(l)} \left( \bar{z}_{l}^{3} - \underline{z}_{l}^{3} \right) \right] \\
  & \boldsymbol S = \int_{z} K_{s} \boldsymbol {\bar{Q}}_{s} \, dz = \sum_{l=1}^{n_{l}} \left[ K_{s} \boldsymbol {\bar{Q}}_{s}^{(l)} \left( \bar{z}_{l} - \underline{z}_{l} \right) \right]
\end{aligned} \right.
\end{equation}
Here, $\bar{\mathbb{Q}}_{m}$ and $\boldsymbol{\bar{Q}}_{s}$ represent the in-plane and out-of-plane stiffness in the global coordinate system, respectively. $K_{s}$ denotes the shear correction factor, while $\bar{z}_{l}$ and $\underline{z}_{l}$ represent the coordinates of the upper and lower ends of the \textit{l}-th layer prism, respectively. The transformation from 3D stiffness in the local coordinate system to 2D stiffness involves a coordinate transformation. Let the 3D stiffness and flexibility matrices in the local coordinate system be denoted by $\boldsymbol{Q}$ and $\boldsymbol{S}$, respectively. The flexibility matrix in the global coordinate system is obtained by transforming the local flexibility matrix as follows:
\begin{equation} \label{eq:16}
\boldsymbol{\bar{S}} = \boldsymbol T  \boldsymbol S \boldsymbol T^{T}
\end{equation}
where the transformation matrix $\boldsymbol T $ is provided in \cite{Ref43}. Then, the 2D stiffness matrix is derived from the 3D flexibility matrix using the following formulation:
\begin{equation} \label{eq:17}
\boldsymbol{\bar{Q}}_{m} = \left[ \begin{matrix}
{\bar{S}}_{11} & {\bar{S}}_{12} & {\bar{S}}_{16} \\
{\bar{S}}_{21} & {\bar{S}}_{22} & {\bar{S}}_{26} \\
{\bar{S}}_{61} & {\bar{S}}_{62} & {\bar{S}}_{66}
\end{matrix} \right]^{-1}, \quad
\boldsymbol{\bar{Q}}_{s} = \left[ \begin{matrix}
{\bar{S}}_{44} & {\bar{S}}_{45} \\
{\bar{S}}_{54} & {\bar{S}}_{55}
\end{matrix} \right]^{-1}
\end{equation}

The plate model shares the same in-plane discretisation as the initial 3D model, i.e., a regular $N_1 \times N_2$ array of pixels. The spatial coordinates of the pixels are given by:

\begin{equation} \label{eq:18}
x_{i} = \left( n_{i} - \frac{1}{2} \right) \frac{L_{i}}{N_{i}}, \quad i = 1, 2, 3 \quad \text{and} \quad n_{i} \in \left[ 1, N_{i} \right]
\end{equation}

Since the Green's operator is calculated in Fourier space, discrete frequencies need to be defined. To reduce the Gibbs phenomenon \cite{Ref44}, a rotation-based sequence is used to define the discrete frequencies in this work \cite{Ref45}:
\begin{equation} \label{eq:19}
\left\{ 
\begin{aligned}
  & \xi_{1} = \frac{N_{1}}{L_{1}} \cdot \sin \left( \bar{\xi}_{1} \right) \cdot \cos \left( \bar{\xi}_{2} \right) \\ 
  & \xi_{2} = \frac{N_{2}}{L_{2}} \cdot \cos \left( \bar{\xi}_{1} \right) \cdot \sin \left( \bar{\xi}_{2} \right) \\ 
\end{aligned}
\right.
\end{equation}
with:
\begin{equation} \label{eq:20}
\bar{\xi}_{i} = \frac{\pi}{N_{i}} 
\left\{ 
\begin{aligned}
  & n_{i} - \frac{N_{i}-1}{2} - 1, \quad \text{if } N_{i} \text{ is odd} \\ 
  & n_{i} - \frac{N_{i}}{2} - 1, \quad \text{if } N_{i} \text{ is even} \\ 
\end{aligned}
\right., \quad i = 1, 2 \quad \text{and} \quad n_{i} \in \left[ 1, N_{i} \right]
\end{equation}

\subsubsection{Offline stage} \label{sec:2.3.2}

In the offline stage, the simulation domain is first partitioned into a finite number of material clusters using the K-means clustering algorithm (Fig.~\ref{fig:F2}). The goal of the clustering process is to ensure that the material points within each cluster exhibit similar mechanical behaviour. A simple and effective way to identify material points with similar mechanical behaviour is to perform multiple DNS calculations on the RVE model to obtain the strain concentration tensor $\mathbb{O}$ for each material point:
\begin{equation} \label{eq:21}
\varepsilon \left( x \right) = \mathbb{O}\left( \boldsymbol{x} \right) : \bar{\varepsilon}
\end{equation}

In this study, the strain concentration tensor is calculated using FFT simulations under six orthogonal loading conditions. After determining the strain concentration tensor, all material points are clustered into \textit{H} material clusters using the K-means algorithm \cite{Ref46}.

The static clustering-based reduced-order model assumes “similar responses once, similar responses always.” This means that material points with initially similar mechanical behaviour are assumed to maintain similar behaviour throughout different loading stages. Consequently, the material clusters defined based on the initial strain concentration tensor remain unchanged. In addition, according to the formula for calculating the interaction tensor, the interaction tensor depends only on the spatial distribution of the material clusters and the properties of the reference medium. Therefore, the interaction tensor can be pre-calculated after clustering and remains constant during the application of specific loads, i.e., the so-called “\textit{offline stage}.”

Since the Green's operators have explicit expressions in Fourier space, the integration by parts for the interaction tensor distribution can be performed efficiently in Fourier space:
\begin{equation} \label{eq:22}
\left\{ \begin{aligned}
  & \int_{\Omega }{{{\chi }^{J}}\left( \boldsymbol x \right){{\mathbb{G}}^{0\left( NN \right)}}\left( \boldsymbol x-\boldsymbol y \right)dy} = {{\mathcal{F}}^{-1}}\left[ {{{\hat{\chi }}}^{J}}\left( \xi  \right):{{{\hat{\mathbb{G}}}}^{0\left( NN \right)}}\left( \xi  \right) \right] \\ 
  & \int_{\Omega }{{{\chi }^{J}}\left( \boldsymbol x \right){{\mathbb{G}}^{0\left( MM \right)}}\left( \boldsymbol x-\boldsymbol y \right)dy} = {{\mathcal{F}}^{-1}}\left[ {{{\hat{\chi }}}^{J}}\left( \xi  \right):{{{\hat{\mathbb{G}}}}^{0\left( MM \right)}}\left( \xi  \right) \right] \\ 
  & \int_{\Omega }{{{\chi }^{J}}\left(\boldsymbol x \right){{\mathcal{G}}^{0\left( MQ \right)}}\left(\boldsymbol x-\boldsymbol y \right)dy} = {{\mathcal{F}}^{-1}}\left[ {{{\hat{\chi }}}^{J}}\left( \xi  \right):{{{\hat{\mathcal{G}}}}^{0\left( MQ \right)}}\left( \xi  \right) \right] \\ 
  & \int_{\Omega }{{{\chi }^{J}}\left( \boldsymbol x \right){{\mathcal{G}}^{0\left( QM \right)}}\left(\boldsymbol x-\boldsymbol y \right)dy} = {{\mathcal{F}}^{-1}}\left[ {{{\hat{\chi }}}^{J}}\left( \xi  \right):{{{\hat{\mathcal{G}}}}^{0\left( QM \right)}}\left( \xi  \right) \right] \\ 
  & \int_{\Omega }{{{\chi }^{J}}\left(\boldsymbol x \right){{G}^{0\left( QQ \right)}}\left(\boldsymbol x-\boldsymbol y \right)dy} = {{\mathcal{F}}^{-1}}\left[ {{{\hat{\chi }}}^{J}}\left( \xi  \right):{{\boldsymbol {\hat{G}}}^{0\left( QQ \right)}}\left( \xi  \right) \right]
\end{aligned} \right.
\end{equation}
The interaction tensor can be calculated using the following equation:
\begin{equation} \label{eq:23}
\left\{ \begin{aligned}
  & {\boldsymbol {D}^{IJ\left( NN \right)}}=\frac{1}{{{c}^{I}}\left| \Omega  \right|}\int_{\Omega }{{{\chi }^{I}}\left(\boldsymbol x \right)\left\{ {{\mathcal{F}}^{-1}}\left[ {{{\hat{\chi }}}^{J}}\left( \xi  \right):{{{\hat{\mathbb{G}}}}^{0\left( NN \right)}}\left( \xi  \right) \right] \right\}dx} \\ 
  & {\boldsymbol {D}^{IJ\left( MM \right)}}=\frac{1}{{{c}^{I}}\left| \Omega  \right|}\int_{\Omega }{{{\chi }^{I}}\left(\boldsymbol x \right)\left\{ {{\mathcal{F}}^{-1}}\left[ {{{\hat{\chi }}}^{J}}\left( \xi  \right):{{{\hat{\mathbb{G}}}}^{0\left( MM \right)}}\left( \xi  \right) \right] \right\}dx} \\ 
  & {\boldsymbol {D}^{IJ\left( MQ \right)}}=\frac{1}{{{c}^{I}}\left| \Omega  \right|}\int_{\Omega }{{{\chi }^{I}}\left(\boldsymbol x \right)\left\{ {{\mathcal{F}}^{-1}}\left[ {{{\hat{\chi }}}^{J}}\left( \xi  \right):{{{\hat{\mathcal{G}}}}^{0\left( MQ \right)}}\left( \xi  \right) \right] \right\}dx} \\ 
  & {\boldsymbol {D}^{IJ\left( QM \right)}}=\frac{1}{{{c}^{I}}\left| \Omega  \right|}\int_{\Omega }{{{\chi }^{I}}\left(\boldsymbol x \right)\left\{ {{\mathcal{F}}^{-1}}\left[ {{{\hat{\chi }}}^{J}}\left( \xi  \right):{{{\hat{\mathcal{G}}}}^{0\left( QM \right)}}\left( \xi  \right) \right] \right\}dx} \\ 
  & {\boldsymbol {D}^{IJ\left( QQ \right)}}=\frac{1}{{{c}^{I}}\left| \Omega  \right|}\int_{\Omega }{{{\chi }^{I}}\left(\boldsymbol x \right)\left\{ {{\mathcal{F}}^{-1}}\left[ {{{\hat{\chi }}}^{J}}\left( \xi  \right):{{{\boldsymbol {\hat{G}}}}^{0\left( QQ \right)}}\left( \xi  \right) \right] \right\}dx} 
\end{aligned} \right.
\end{equation}
Similar to the FoPT, the calculation of the interaction tensor for the CPT is detailed in Appendix~\ref{sec:A.2}.

Additionally, in the case of damage or nonlinear behaviour at local material points within the model, the properties of the reference medium can be adjusted to improve convergence, e.g., by using a self-consistent update scheme. However, the properties of the reference medium directly affect the calculation of the Green's operators, which in turn influences the calculation of the interaction tensor. According to Eq.~(\ref{eq:23}), computing the interaction tensor in its discrete form is computationally expensive. Frequent re-calculation of the interaction tensor would result in significant computational cost. A feasible approach is to decouple the parameters associated with the reference medium from the discrete frequencies. This allows the computationally intensive tasks related to the discrete frequency terms to be performed in the offline stage, while the online computations focus on updating the material coefficients corresponding to the reference medium. As a result, the Green's operators and the interaction tensor can be updated efficiently. The separable formulation for the CPT is provided in Appendix~\ref{sec:A.2}.

The separation form of the Green's operators for the FoPT can be expressed as:
\begin{equation} \label{eq:24}
\left\{ \begin{aligned}
  & \hat{G}_{khij}^{0(NN)}=\frac{1}{4\mu _{m}^{0}}\hat{G}_{khij}^{1(NN)}+\frac{\lambda _{m}^{0}+\mu _{m}^{0}}{\mu _{m}^{0}\left( \lambda _{m}^{0}+2\mu _{m}^{0} \right)}\hat{G}_{khij}^{2(NN)} \\ 
  & \hat{G}_{khij}^{0(MM)}=\frac{1}{\mu _{b}^{0}{{\left| \xi  \right|}^{2}}+{{G}^{0}}}\hat{G}_{khij}^{1(MM)}+\frac{1}{\left( \lambda _{b}^{0}+2\mu _{b}^{0} \right)}\hat{G}_{khij}^{2(MM)} \\ 
  & \hat{G}_{khi}^{0(MQ)}=\frac{1}{\mu _{b}^{0}{{\left| \xi  \right|}^{2}}+{{G}^{0}}}\hat{G}_{khi}^{1(MQ)} \\ 
  & \hat{G}_{kij}^{0(QM)}=\frac{1}{\mu _{b}^{0}{{\left| \xi  \right|}^{2}}+{{G}^{0}}}\hat{G}_{kij}^{1(QM)} \\ 
  & \hat{G}_{ki}^{0(QQ)}=\frac{1}{\mu _{b}^{0}{{\left| \xi  \right|}^{2}}+{{G}^{0}}}\hat{G}_{ki}^{1(QQ)}+\frac{1}{{{G}^{0}}}\hat{G}_{ki}^{2(QQ)} 
\end{aligned} \right.
\end{equation}
where:
\begin{equation} \label{eq:25}
\left\{ \begin{aligned}
  & \hat{G}_{khij}^{1(NN)}=\frac{{{\delta }_{ik}}{{\xi }_{j}}{{\xi }_{l}}+{{\delta }_{il}}{{\xi }_{j}}{{\xi }_{k}}+{{\delta }_{jl}}{{\xi }_{i}}{{\xi }_{k}}+{{\delta }_{jk}}{{\xi }_{i}}{{\xi }_{l}}}{{{\left| \xi  \right|}^{2}}}\hat{G}_{khij}^{2(NN)}=-\frac{{{\xi }_{i}}{{\xi }_{j}}{{\xi }_{k}}{{\xi }_{l}}}{{{\left| \xi  \right|}^{4}}} \\ 
  & \hat{G}_{khij}^{1(MM)}=\frac{{{\delta }_{ih}}{{\xi }_{j}}{{\xi }_{k}}+{{\delta }_{ik}}{{\xi }_{h}}{{\xi }_{j}}}{2}-\frac{{{\xi }_{h}}{{\xi }_{i}}{{\xi }_{j}}{{\xi }_{k}}}{{{\left| \xi  \right|}^{2}}}\hat{G}_{khij}^{2(MM)}=\frac{{{\xi }_{h}}{{\xi }_{i}}{{\xi }_{j}}{{\xi }_{k}}}{{{\left| \xi  \right|}^{4}}} \\ 
  & \hat{G}_{khi}^{1(MQ)}=\frac{i{{\delta }_{ih}}{{\xi }_{k}}+{{\delta }_{ik}}{{\xi }_{h}}}{2}-\frac{i{{\xi }_{h}}{{\xi }_{i}}{{\xi }_{k}}}{{{\left| \xi  \right|}^{2}}} \\ 
  & \hat{G}_{kij}^{1(QM)}=\frac{i{{\xi }_{i}}{{\xi }_{j}}{{\xi }_{k}}}{{{\left| \xi  \right|}^{2}}}-i{{\delta }_{ik}}{{\xi }_{j}} \\ 
  & \hat{G}_{ki}^{1(QQ)}={{\delta }_{ik}}-\frac{{{\xi }_{i}}{{\xi }_{k}}}{{{\left| \xi  \right|}^{2}}},\quad \hat{G}_{ki}^{2(QQ)}=\frac{{{\xi }_{i}}{{\xi }_{k}}}{{{\left| \xi  \right|}^{2}}} 
\end{aligned} \right.
\end{equation}

For the operators $\hat{G}_{khij}^{0(NN)}$, $\hat{G}_{khij}^{0(MM)}$, $\hat{G}_{khi}^{0(MQ)}$, $\hat{G}_{kij}^{0(QM)}$, and $\hat{G}_{ki}^{0(QQ)}$, the coefficient $\rho = \frac{1}{\mu _{b}^{0}{{\left| \xi  \right|}^{2}}+{{G}^{0}}}$ cannot be directly decoupled into discrete frequencies and the material parameters of the reference medium. To enable efficient re-calculation of the Green's operators when the material parameters of the reference medium are varied, the coefficient $\rho$ is assumed to satisfy the following condition:
\begin{equation} \label{eq:26}
\frac{{{\rho }^{i+1}}}{{{\rho }^{i}}} = C
\end{equation}
where ${\rho}^{i}$ and ${\rho}^{i+1}$ represent the coefficients before and after the update, respectively, and $C$ is a constant. Under this assumption, the updated coefficient can be obtained by multiplying the previous coefficient by $C$. If this assumption is satisfied, the Lamé constants of the reference medium must meet the following condition:
\begin{equation} \label{eq:27}
\frac{{{\left( \mu _{b}^{0}{{\left| \xi  \right|}^{2}}+{{G}^{0}} \right)}^{i}}}{{{\left( \mu _{b}^{0}{{\left| \xi  \right|}^{2}}+{{G}^{0}} \right)}^{i+1}}} = C \Leftrightarrow \frac{{{\left( \mu _{b}^{0} \right)}^{i}}}{{{\left( \mu _{b}^{0} \right)}^{i+1}}} = \frac{{{\left( {{G}^{0}} \right)}^{i}}}{{{\left( {{G}^{0}} \right)}^{i+1}}} = C
\end{equation}
This means that the ratio of the Lamé constant ($\mu _{b}^{0}$) before and after the update is equal to the ratio of ${G}^{0}$ before and after the update. However, this assumption is not universally valid when the material parameters of the reference medium change. As discussed in Section 6.1 of Li et al.'s \cite{Ref36} work on solving the FoPT using FFT-based methods, changes in ${G}^{0}$ show strong insensitivity to the convergence of FFT methods. If ${G}^{0}$ exhibits similar insensitivity in cluster-based computations, then the Green's operators can be efficiently recalculated based on the above assumptions. Further discussion on the sensitivity of ${G}^{0}$ is provided in Sec.~\ref{sec:4.1}.

The numerical implementation of the offline stage is summarised in Algorithm~\ref{al:offline}.
\RestyleAlgo{boxed}
\begin{algorithm}
\DontPrintSemicolon
\SetAlgoLined
 \textbf{Data}: $\mathbb{A},\mathbb{B},\mathbb{D},\boldsymbol{S},\mathbb{A}^0,\mathbb{D}^0,\boldsymbol{S}^0 ,\mathbb{G}^{0\left( NN \right)}, \mathbb{G}^{0\left( MM \right)},\mathcal{G}^{0\left( MQ \right)},\mathcal{G}^{0\left( QM \right)},\boldsymbol G^{0\left( QQ \right)}$\\
 \smallskip
\it{ (a) Strain concentration tensor obtained from FFT calculation}\\
\For {$l=1:8$} {
(i)  ${\boldsymbol \varepsilon}^l(\boldsymbol{x}),{\boldsymbol \phi}^l(\boldsymbol{x}),{\boldsymbol \gamma}^l(\boldsymbol{x})$ calculated by FFT-solver\\
(ii)  ${{\mathbb{O}}^{l}}\left(\boldsymbol{x} \right)={{\left[ \text{ }{\boldsymbol \varepsilon}^l(\boldsymbol{x}),{\boldsymbol \phi}^l(\boldsymbol{x}),{\boldsymbol \gamma}^l(\boldsymbol{x}) \right]}^{T}}$
} 
\smallskip
\it{ (b) k-means clustering: H clusters}\\
\smallskip
\it{ (c) Calculate the interaction tensor}\\
\For {$J=1:H$} {
(i)  ${{\hat{\chi }}^{J}}\left( \boldsymbol{\xi}  \right)=\mathcal{F}\left[ {{\chi }^{J}}\left(\boldsymbol{x} \right) \right]$\\
(ii)  Calculate the integral in Fourier space based on Eq. ( \ref{eq:22} )\\
\For {$I=1:H$} {
(iii)  Calculate the volume fraction: $c^I$\\
(iv)  Calculate the integral in real space based on Eq. ( \ref{eq:23} )\\
}
} 
\rm\textbf{Result}: 
$\boldsymbol{D}^{IJ\left( \sim,\sim \right)}$\\
\caption{Numerical implementation of the offline stage.}\label{al:offline}
\end{algorithm} 

\subsubsection{Online stage} \label{sec:2.3.3}

When the average strain increment is specified, the discrete incremental Lippmann–Schwinger equations, Eq.~(\ref{eq:13}), must be solved online. This process involves updating the constitutive behaviour at local material points based on the calculated response, i.e., the so-called "\textit{online stage}". Equation~(\ref{eq:13}) can be solved using the Newton–Raphson method, where the residual of the equations is expressed as:
\begin{equation} \label{eq:28}
\left\{ \begin{aligned}
  & {\boldsymbol {r}^{I\left( N \right)}} = \Delta {\boldsymbol {\varepsilon }^{I}} + \sum\limits_{J=1}^{H}{{\boldsymbol {D}^{IJ\left( NN \right)}}\left[ \Delta {\boldsymbol {N }^{J}} - {{\mathbb{A}}^{0}} : \Delta {\boldsymbol {\varepsilon }^{J}} \right]} - \Delta \bar{\boldsymbol \varepsilon} \\
  & {\boldsymbol {r}^{I\left( M \right)}} = \Delta {\boldsymbol {\phi }^{I}} + \sum\limits_{J=1}^{H}{{\boldsymbol {D}^{IJ\left( MM \right)}}\left[ \Delta {\boldsymbol {M }^{J}} - {{\mathbb{D}}^{0}} : \Delta {\boldsymbol {\phi }^{J}} \right]} + \sum\limits_{J=1}^{H}{{\boldsymbol {D}^{IJ\left( MQ \right)}}\left[ \Delta {\boldsymbol {Q}^{J}} - {\boldsymbol {S}^{0}} \cdot \Delta {\boldsymbol {\gamma }^{J}} \right]} - \Delta \bar{\boldsymbol \phi} \\
  & {\boldsymbol {r}^{I\left( Q \right)}} = \Delta {\boldsymbol {\gamma }^{I}} + \sum\limits_{J=1}^{H}{{\boldsymbol {D}^{IJ\left( QM \right)}}\left[ \Delta {\boldsymbol {M }^{J}} - {{\mathbb{D}}^{0}} : \Delta {{\boldsymbol \phi }^{J}} \right]} + \sum\limits_{J=1}^{H}{{\boldsymbol {D}^{IJ\left( QQ \right)}}\left[ \Delta {\boldsymbol {Q}^{J}} - {\boldsymbol {S}^{0}} \cdot \Delta {{\boldsymbol \gamma }^{J}} \right]} - \Delta \bar{\boldsymbol \gamma}
\end{aligned} 
\right.
\end{equation}

When the Newton–Raphson iterative method is used to solve the system of equations, the Jacobian matrix must be computed to improve convergence. For the FoPT, the problem is divided into three systems of equations, requiring the calculation of three residuals and their corresponding Jacobian matrices:
\begin{equation} \label{eq:29}
\left\{ \begin{aligned}
  & {\boldsymbol {J}^{\left( N \right)}} = \frac{\partial {{\boldsymbol r}^{\left( N \right)}}\left( \Delta \boldsymbol{\varepsilon}  \right)}{\partial \Delta \boldsymbol{\varepsilon}} \\
  & {\boldsymbol {J}^{\left( M \right)}} = \frac{\partial {{\boldsymbol r}^{\left( M \right)}}\left( \Delta \boldsymbol{\phi}  \right)}{\partial \Delta \boldsymbol{\phi}} \\
  & {\boldsymbol {J}^{\left( Q \right)}} = \frac{\partial {{\boldsymbol r}^{\left( Q \right)}}\left( \Delta \boldsymbol{\gamma}  \right)}{\partial \Delta \boldsymbol{\gamma}}
\end{aligned} 
\right.
\end{equation}
Substituting Eq. ( \ref{eq:28} ) into Eq. ( \ref{eq:29} ) yields:
\begin{equation} \label{eq:30}
\left\{ \begin{aligned}
  & {\boldsymbol {J}^{\left( N \right)}} = {\boldsymbol {D}^{IJ\left( NN \right)}} \frac{\partial \Delta \boldsymbol{N}}{\partial \Delta \boldsymbol{\varepsilon}} + \mathrm{\boldsymbol I} - {\boldsymbol {D}^{IJ\left( NN \right)}} \mathbb{A} \\
  & {\boldsymbol {J}^{\left( M \right)}} = {\boldsymbol {D}^{IJ\left( MM \right)}} \frac{\partial \Delta \boldsymbol{M}}{\partial \Delta \boldsymbol{\phi}} + \mathrm{\boldsymbol I} - {\boldsymbol {D}^{IJ\left( MM \right)}} \mathbb{D} \\
  & {\boldsymbol {J}^{\left( Q \right)}} = {\boldsymbol {D}^{IJ\left( QQ \right)}} \frac{\partial \Delta \boldsymbol{Q}}{\partial \Delta \boldsymbol{\gamma}} + \mathrm{\boldsymbol I} - {\boldsymbol {D}^{IJ\left( QQ \right)}} \boldsymbol S
\end{aligned}
\right.
\end{equation}
where,
\begin{equation} \label{eq:31}
\mathbb{A} = \left[ \begin{matrix}
   {{\mathbb{A}}^{0}} & {} & 0  \\
   {} & \ddots  & {}  \\
   0 & {} & {{\mathbb{A}}^{0}}  \\
\end{matrix} \right],\quad
\mathbb{D} = \left[ \begin{matrix}
   {{\mathbb{D}}^{0}} & {} & 0  \\
   {} & \ddots  & {}  \\
   0 & {} & {{\mathbb{D}}^{0}}  \\
\end{matrix} \right],\quad
\boldsymbol S = \left[ \begin{matrix}
   {\boldsymbol {S}^{0}} & {} & 0  \\
   {} & \ddots  & {}  \\
   0 & {} & {\boldsymbol {S}^{0}}  \\
\end{matrix} \right]
\end{equation}

Although the Newton–Raphson iterative method can be employed to solve the system of equations, for continuous Lippmann–Schwinger equations the solution is independent of the choice of reference material, which only affects the convergence speed. However, for cluster-based reduced-order models, the discretisation can be interpreted as a piecewise uniform assumption of material clusters, where the material points in the RVE model cannot strictly satisfy equilibrium conditions. Therefore, the solution of such reduced-order models depends on the choice of the reference medium. Although this error can be mitigated by increasing the number of material clusters, the increase in degrees of freedom consequently raises the computational cost. To balance computational efficiency and accuracy, a self-consistent scheme can be used to update the reference medium, which is also the purpose of separating the reference material parameters and the discrete frequency terms in the Green's operator, as discussed in Sec.~\ref{sec:2.3.2}.

In the self-consistent scheme, the stiffness tensor of the reference material should approximate the tangent stiffness of the RVE model:
\begin{equation} \label{eq:32}
{{\bar{\mathbb{A}}^{0}}} \to \bar{\mathbb{A}},\quad {{\bar{\mathbb{D}}^{0}}} \to \bar{\mathbb{D}},\quad {{\bar{\boldsymbol S}}^{0}} \to \bar{\boldsymbol S}
\end{equation}

Li et al. \cite{Ref36} determined the Lamé constants of the stiffness based on the eigenvalues of the stiffness tensor. However, in the clustering-based model, this approach results in lower accuracy. To achieve a more accurate approximation of the isotropic stiffness, this study adopts a projection-based self-consistent scheme with improved accuracy \cite{Ref28}. The implementation details of this scheme are provided in Appendix~\ref{sec:A.3}. The update of ${{\bar{\boldsymbol S}}^{0}}$ is governed by ${G}^{0}$ and is only defined for the FoPT. The value of ${G}^{0}$ is determined according to the assumption described in Eq.~(\ref{eq:27}). Additionally, during the offline stage of the FFT calculations, the Lamé constants of the reference material are determined following the procedure described by \cite{Ref36}.

The numerical implementation of the online stage is summarised in Algorithm~\ref{al:online}.

\RestyleAlgo{boxed}
\begin{algorithm}
\DontPrintSemicolon
\SetAlgoLined
 \textbf{Data}: $\mathbb{A},\mathbb{B},\mathbb{D},\boldsymbol{S},\mathbb{A}^0,\mathbb{D}^0,\boldsymbol{S}^0 ,\mathbb{G}^{0\left( NN \right)}, \mathbb{G}^{0\left( MM \right)},\mathcal{G}^{0\left( MQ \right)},\mathcal{G}^{0\left( QM \right)},\boldsymbol G^{0\left( QQ \right)},\boldsymbol{D}^{IJ\left( \sim,\sim \right)}$\\
 \smallskip
\it{ (a) Input: $\Delta\bar{\boldsymbol \varepsilon},\Delta\bar{\boldsymbol \phi},\Delta\bar{\boldsymbol \gamma}$}\\
 \smallskip
\it{ (b) Initialization: $\Delta{\boldsymbol \varepsilon}^0(\bm x)=\Delta\bar{\boldsymbol \varepsilon},\Delta{\boldsymbol \phi}^0(\bm x)=\Delta\bar{\boldsymbol \phi},\Delta{\boldsymbol \gamma}^0(\bm x)=\Delta\bar{\boldsymbol \gamma},Err=+\infty$}\\
\smallskip
\it{ (c) Newton-Raphson iterations}\\
\smallskip
\While {$Err>Err_{tol}$} {
(i)  Calculate the stress increment: $\left\{  
\begin{aligned}  
  & \Delta \bm{N}^{I,i+1} (\bm x) = \mathbb{A}^{I} (\bm x) : \Delta 
 \bm{\varepsilon}^{I,i+1} (\bm x) + \mathbb{B}^{I} (\bm x) : \Delta  \bm{\phi}^{I,i+1} (\bm x) \\  
  & \Delta  \bm{M}^{I,i+1} (\bm x) = \mathbb{B}^{I} (\bm x) : \Delta 
 \bm{\varepsilon}^{I,i+1} (\bm x) + \mathbb{D}^{I} (\bm x) : \Delta  \bm{\phi}^{I,i+1} (\bm x) \\  
  & \Delta  \bm{Q}^{I,i+1} (\bm x) = \bm{S}^{I} (\bm x) \cdot \Delta 
 \bm{\gamma}^{I,i+1} (\bm x)  
\end{aligned}  
\right.$
\\
(ii) Calculate the residual ($\bm{r}^{I(N),i},\bm{r}^{I(M),i},\bm{r}^{I(Q),i}$) by Eq. ( \ref{eq:28} )\\
(iii) Calculate the Jacobian matrix ($\bm{J}^{I(N),i},\bm{J}^{I(M),i},\bm{J}^{I(Q),i}$) by Eq. ( \ref{eq:30} )\\
(iv) Update the strain increment: $\left\{
\begin{aligned}
  & \Delta \bm{\varepsilon}^{I,i+1} = \Delta \bm{\varepsilon}^{I,i} - \bm{J}^{I(N),i} \backslash \bm{r}^{I(N),i} \\  
  & \Delta \bm{\phi}^{I,i+1} = \Delta \bm{\phi}^{I,i} - \bm{J}^{I(M),i} \backslash \bm{r}^{I(M),i} \\  
  & \Delta \bm{\gamma}^{I,i+1} = \Delta \bm{\gamma}^{I,i} - \bm{J}^{I(Q),i} \backslash \bm{r}^{I(Q),i}  
\end{aligned}
\right.$
\\
(v) Calculate error
}  
\smallskip
\it{ (d) Update the reference medium by Appendix   \ref{sec:A.3}}\\
\smallskip
\rm\textbf{Result}: 
$\Delta  \bm{\varepsilon}(\bm x),\Delta  \bm{\phi}(\bm x),\Delta  \bm{\gamma}(\bm x),\Delta \bm{N}(\bm x),\Delta  \bm{M}(\bm x),\Delta  \bm{Q}(\bm x)$\\
\caption{Numerical implementation of the online stage.}\label{al:online}
\end{algorithm} 

\subsubsection{Non-linear analysis} \label{sec:2.3.4}

Two critical challenges emerge in the implementation of the two-stage "offline–online" framework for non-linear analysis:

(i) Error propagation from linear clustering inheritance: Directly transferring the initial linear clustering configuration to the non-linear regime introduces cumulative approximation errors. However, previous studies have shown that the initial clustering distribution retains reasonable accuracy, and for strongly non-linear behaviour, refinement strategies such as sub-clustering can improve accuracy — an aspect beyond the scope of this study.

(ii) Constitutive modelling limitations in plate formulations: Traditional plate models inherently compromise out-of-plane material characterisation through kinematic assumptions, rendering them inadequate for capturing true non-linear constitutive responses in heterogeneous laminates. To address this issue, we implement a nested model order reduction scheme in which:
\begin{itemize}
\item Material nonlinearities (plasticity/progressive damage) are resolved at 3D material points.
\item Updated stress states are consistently projected back to the plate model.
\end{itemize}

The strain ($\bm\varepsilon_s, \bm\gamma_s$) at a given position in the solid model can be determined from the strain $\bm\varepsilon$ and curvature $\bm\phi$ of the plate model, along with the distance $z$ from the target position to the mid-plane \cite{Ref42}:
\begin{equation} \label{eq:33}
\left\{ \begin{aligned}
  & \bm{\varepsilon}_s = \bm{\varepsilon}+z  \cdot \bm{\phi}\\
  & \bm{\gamma}_s = \bm\gamma
\end{aligned} 
\right.
\end{equation}

Furthermore, we employ an incremental solution strategy for non-linear analysis, where macroscopic loading is applied as a homogeneous strain to the RVE. The RVE stress field is calculated through the SCA, followed by homogenisation to obtain the effective stress resultants ($\bar{\bm{N}}, \bar{\bm{M}}, \bar{\bm{Q}}$). These homogenised quantities are derived through volume averaging of the stress and moment resultant distributions of the RVE:
\begin{equation} \label{eq:34}
\left\{ \begin{aligned}
  & \bar{\bm{N}} = \sum\limits_{I=1}^{H}{{{\chi }^{I}} \bm N^{I}(\bm x)} \\  
  & \bar{\bm{M}} = \sum\limits_{I=1}^{H}{{{\chi }^{I}} \bm M^{I}(\bm x)} \\  
  & \bar{\bm{Q}} = \sum\limits_{I=1}^{H}{{{\chi }^{I}} \bm Q^{I}(\bm x)} \\
\end{aligned} 
\right.
\end{equation}

In such cases, the constitutive laws may depend on both strain and strain rates in specific non-linear models. The computational framework for non-linear analysis is summarised in Algorithm~\ref{al:nonlinear}.

\RestyleAlgo{boxed}
\begin{algorithm}
\DontPrintSemicolon
\SetAlgoLined
 \textbf{Data}: $\mathbb{A},\mathbb{B},\mathbb{D},\boldsymbol{S}$ relating to $\bm \varepsilon(\bm x),\bm \phi(\bm x),\bm \gamma(\bm x),\dot {\bm \varepsilon}(\bm x),\dot{\bm \phi}(\bm x),\dot{\bm \gamma}(\bm x)$\\
 \smallskip
\rm\textbf{begin} \\
\For {$\bm{I}=1:\bm{I}_T$} {
(i)  Time step updating:\\ ${\bar {\bm \varepsilon}}_I(t)=\bar{\bm \varepsilon}(t_I),\bar{\bm \phi}_I(t)=\bar{\bm \phi}(t_I),\bar{\bm \gamma}_I(t)=\bar{\bm \gamma}(t_I)$\\
(ii) Calculate the strain and stress distributions in the current step by Algorithm \ref{al:online}:\\ $\bm \varepsilon_I(\bm x),\bm \phi_I(\bm x),\bm \gamma_I(\bm x),\bm N_I(\bm x),\bm M_I(\bm x),\bm Q_I(\bm x)$\\
(iii) Calculate the effective stress resultants by Eq. ( \ref{eq:34} ):\\
$\bar {\bm N_I}(\bm x),\bar{\bm M_I}(\bm x),\bar{\bm Q_I}(\bm x)$\\
(iv) Update the $\mathbb{A}_{I+1},\mathbb{B}_{I+1},\mathbb{D}_{I+1},\boldsymbol{S}_{I+1}$ according to $\bm \varepsilon_I(\bm x),\bm \phi_I(\bm x),\bm \gamma_I(\bm x),\dot {\bm \varepsilon_I}(\bm x),\dot{\bm \phi_I}(\bm x),\dot{\bm \gamma_I}(\bm x)$
}  

\textbf{end}\\
\smallskip
\rm\textbf{Result}: 
$\bar {\bm N},\bar{\bm M},\bar{\bm Q}$ \\
\caption{Computational framework of non-linear analysis.}\label{al:nonlinear}
\end{algorithm} 

\section{Results} \label{sec:Results}

Linear analyses of periodic perforated plates (both symmetric and asymmetric) and a twill woven composite are performed, comparing the proposed method with the DNS method (FFT-based approach). Subsequently, non-linear analyses are carried out on an elasto-plastic periodic perforated plate and a practical plain woven composite. For the latter, the predictions obtained using the proposed method are compared with experimental results. The computation time comparison for all examples is discussed in Sec.~\ref{sec:4.2}.

\subsection{Linear analysis of periodic perforated plate}\label{sec:3.1}

\begin{figure}[htbp]
\centering
\includegraphics[width=16cm]{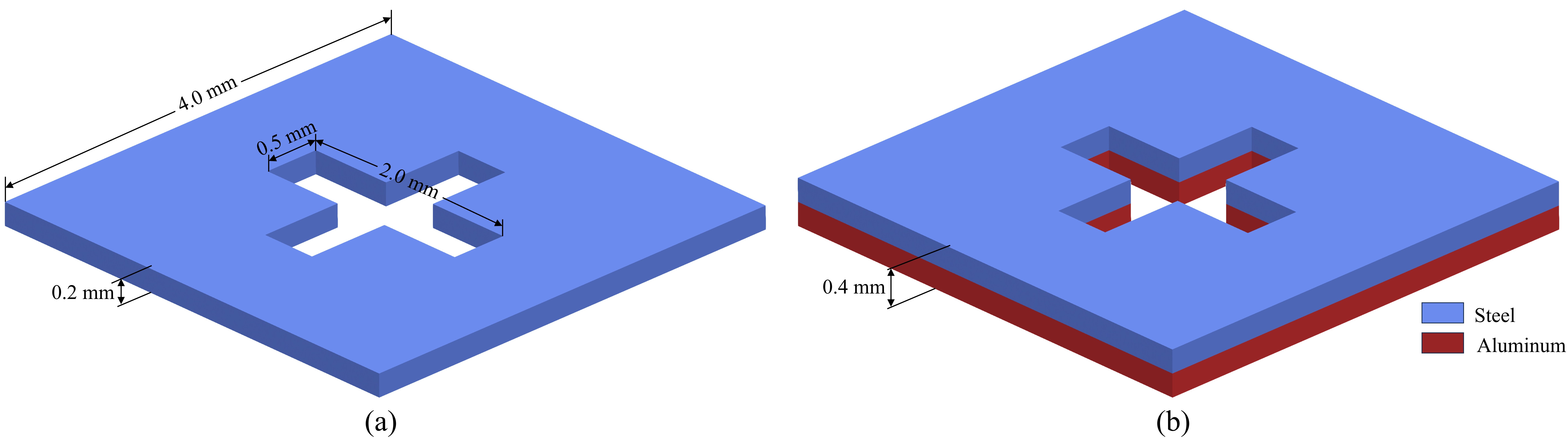}
\caption{\em{Geometric details of the periodic perforated plates. (a) symmetric plate; (b) asymmetric plate.}}
\label{fig:F3}
\end{figure}

The periodic perforated plates considered in this example feature “+”-shaped pores, with the symmetric plate composed of steel and the asymmetric plate consisting of an aluminium–steel laminate. The dimensions and geometric details of the RVE models are shown in Fig.~\ref{fig:F3}. The Young’s modulus and Poisson’s ratio of the steel plate are 200 GPa and 0.3, respectively \cite{Ref47}, while those of the aluminium plate are 70 GPa and 0.33, respectively \cite{Ref48}. Both models have an in-plane resolution of $320 \times 320$, and a convergence study of the selected resolution is presented in \cite{Ref36}.

The linear analysis is performed to determine the effective properties of the RVE in the elastic state and requires six orthogonal loading conditions. Average in-plane strain is applied to the RVE model to calculate the effective membrane properties of the periodic plate. The three cases considered are as follows:
\begin{itemize}
\item Case. 1: $\bar{\bm \varepsilon}=\left[ \begin{matrix}   0.01 & 0 \\
   0 & 0 \end{matrix} \right],\bar{\bm \phi}=\bm{0}$
\item Case. 2: $\bar{\bm \varepsilon}=\left[ \begin{matrix}   0 & 0 \\
   0 & 0.01 \end{matrix} \right],\bar{\bm \phi}=\bm{0}$
\item Case. 3: $\bar{\bm \varepsilon}=\left[ \begin{matrix}   0 & 0.01 \\
   0.01 & 0 \end{matrix} \right],\bar{\bm \phi}=\bm{0}$
\end{itemize}

The effective bending properties are determined from the following three loading cases:
\begin{itemize}
\item Case. 4: $\bar{\bm \varepsilon}=\bm{0},\bar{\bm \phi}=\left[ \begin{matrix}   0.01 & 0 \\  0 & 0 \end{matrix} \right]$
\item Case. 5: $\bar{\bm \varepsilon}=\bm{0},\bar{\bm \phi}=\left[ \begin{matrix}   0 & 0 \\  0 & 0.01 \end{matrix} \right]$
\item Case. 6: $\bar{\bm \varepsilon}=\bm{0},\bar{\bm \phi}=\left[ \begin{matrix}   0 & 0.01 \\  0.01 & 0 \end{matrix} \right]$
\end{itemize}

The six load cases are analysed using both the FFT method and the SCA approach. In the SCA calculations, five different cluster numbers $H$ (i.e., 32, 64, 128, 256, and 512) are considered. In addition, both the CPT and FoPT problems are analysed. The homogenised properties of symmetric and asymmetric perforated plates are presented in Tables~\ref{tab:T1} and \ref{tab:T2}, respectively. Homogenised properties with zero values (e.g., $B_{ij} = 0$ for the symmetric perforated plate) and symmetric properties (e.g., $A_{11} = A_{22}$) are omitted. Since the thickness-to-length ratio of all plates does not exceed 0.1, the homogenised properties obtained from CPT and FoPT exhibit minimal differences. However, as the thickness-to-length ratio increases, the discrepancy in $D_{ij}$ also increases. Furthermore, a comparison between the SCA and FFT results indicates that as $H$ increases, the homogenised properties progressively converge to those obtained by FFT. When $H = 128$, the error remains within 1\%. This demonstrates that the SCA can achieve high accuracy in predicting homogenised properties using a limited number of clusters.

\begin{table}[h]
    \renewcommand*{\arraystretch}{1.5}
    \centering
    \caption{Homogenized properties of the symmetric plate ($A_{ij}$ [MPa$\cdot$m], $D_{ij}$ [N$\cdot$mm]).}
    \label{tab:T1}      
    \resizebox{\textwidth}{!}{ 
    \begin{tabular}{l c c c c c c c c c c c c}
        \hline
        & \multicolumn{2}{c}{$H=32$} & \multicolumn{2}{c}{$H=64$} & \multicolumn{2}{c}{$H=128$} & \multicolumn{2}{c}{$H=256$} & \multicolumn{2}{c}{$H=512$} & \multicolumn{2}{c}{$\text{FFT}$} \\
        & CPT & FoPT & CPT & FoPT & CPT & FoPT & CPT & FoPT & CPT & FoPT & CPT & FoPT \\
        \hline
        $A_{11}$ & 27.760 & 28.144 & 27.648 & 27.792 & 27.342 & 27.390 & 27.272 & 27.312 & 27.215 & 27.212 & 27.127 & 27.127 \\
        $A_{12}$ & 7.636 & 7.543 & 7.745 & 7.494 & 7.634 & 7.608 & 7.637 & 7.619 & 7.634 & 7.622 & 7.629 & 7.629 \\
        $A_{33}$& 8.893 & 8.817 & 8.988 & 8.837 & 8.916 & 8.868 & 8.911 & 8.873 & 8.908 & 8.891 & 8.890 & 8.890 \\
        \hline
        $D_{11}$ & 105.254 & 104.486 & 105.189 & 103.441 & 104.539 & 102.835 & 104.306 & 102.473 & 104.066 & 101.923 & 103.715 & 101.320 \\
        $D_{12}$ & 21.649 & 21.262 & 21.551 & 20.963 & 21.231 & 20.712 & 21.203 & 20.507 & 21.170 & 20.405 & 21.115 & 20.483 \\
        $D_{33}$ & 44.713 & 43.373 & 44.529 & 43.080 & 44.060 & 42.766 & 43.782 & 42.338 & 43.442 & 41.969 & 41.974 & 40.857 \\
        \hline
    \end{tabular}
    }
\end{table}

\begin{table}[h]
    \renewcommand*{\arraystretch}{1.5}
    \centering
    \caption{Homogenized properties of the asymmetric plate ($A_{ij}$ [MPa$\cdot$m], $B_{ij}$ [kN], $D_{ij}$ [N$\cdot$mm]).}
    \label{tab:T2}      
    \resizebox{\textwidth}{!}{ 
    \begin{tabular}{l c c c c c c c c c c c c}
        \hline
        & \multicolumn{2}{c}{$H=32$} & \multicolumn{2}{c}{$H=64$} & \multicolumn{2}{c}{$H=128$} & \multicolumn{2}{c}{$H=256$} & \multicolumn{2}{c}{$H=512$} & \multicolumn{2}{c}{$\text{FFT}$} \\
        & CPT & FoPT & CPT & FoPT & CPT & FoPT & CPT & FoPT & CPT & FoPT & CPT & FoPT \\
        \hline
        $A_{11}$ & 37.564 & 37.710 & 37.396 & 37.153 & 36.980 & 36.948 & 36.937 & 36.943 & 36.834 & 36.833 & 36.737 & 36.737 \\
        $A_{12}$ & 10.518 & 10.307 & 10.408 & 10.338 & 10.536 & 10.477 & 10.526 & 10.495 & 10.528 & 10.503 & 10.512 & 10.512 \\
        $A_{33}$ & 11.923 & 11.732 & 12.017 & 11.806 & 11.963 & 11.856 & 11.984 & 11.917 & 11.983 & 11.939 & 11.961 & 11.961 \\
        \hline
        $B_{11}$ & 1.751 & 1.729 & 1.753 & 1.736 & 1.750 & 1.741 & 1.753 & 1.747 & 1.753 & 1.749 & 1.752 & 1.750 \\
        $B_{12}$ & 0.436 & 0.422 & 0.447 & 0.442 & 0.463 & 0.457 & 0.466 & 0.462 & 0.469 & 0.467 & 0.471 & 0.470 \\
        $B_{33}$ & 0.580 & 0.561 & 0.585 & 0.566 & 0.583 & 0.572 & 0.585 & 0.578 & 0.585 & 0.581 & 0.584 & 0.580 \\
        \hline
        $D_{11}$ & 556.817 & 543.396 & 556.945 & 540.198 & 553.332 & 537.799 & 551.954 & 536.508 & 550.563 & 533.935 & 548.522 & 531.358 \\
        $D_{12}$ & 124.165 & 121.924 & 125.059 & 120.129 & 122.797 & 118.647 & 122.559 & 118.432 & 122.269 & 117.918 & 122.051 & 117.618 \\
        $D_{33}$ & 225.397 & 214.526 & 225.389 & 214.012 & 222.559 & 213.086 & 221.852 & 212.145 & 220.070 & 210.259 & 212.592 & 204.360 \\
        \hline
    \end{tabular}
    }
\end{table}

\begin{figure}[htbp]
\centering
\includegraphics[width=16cm]{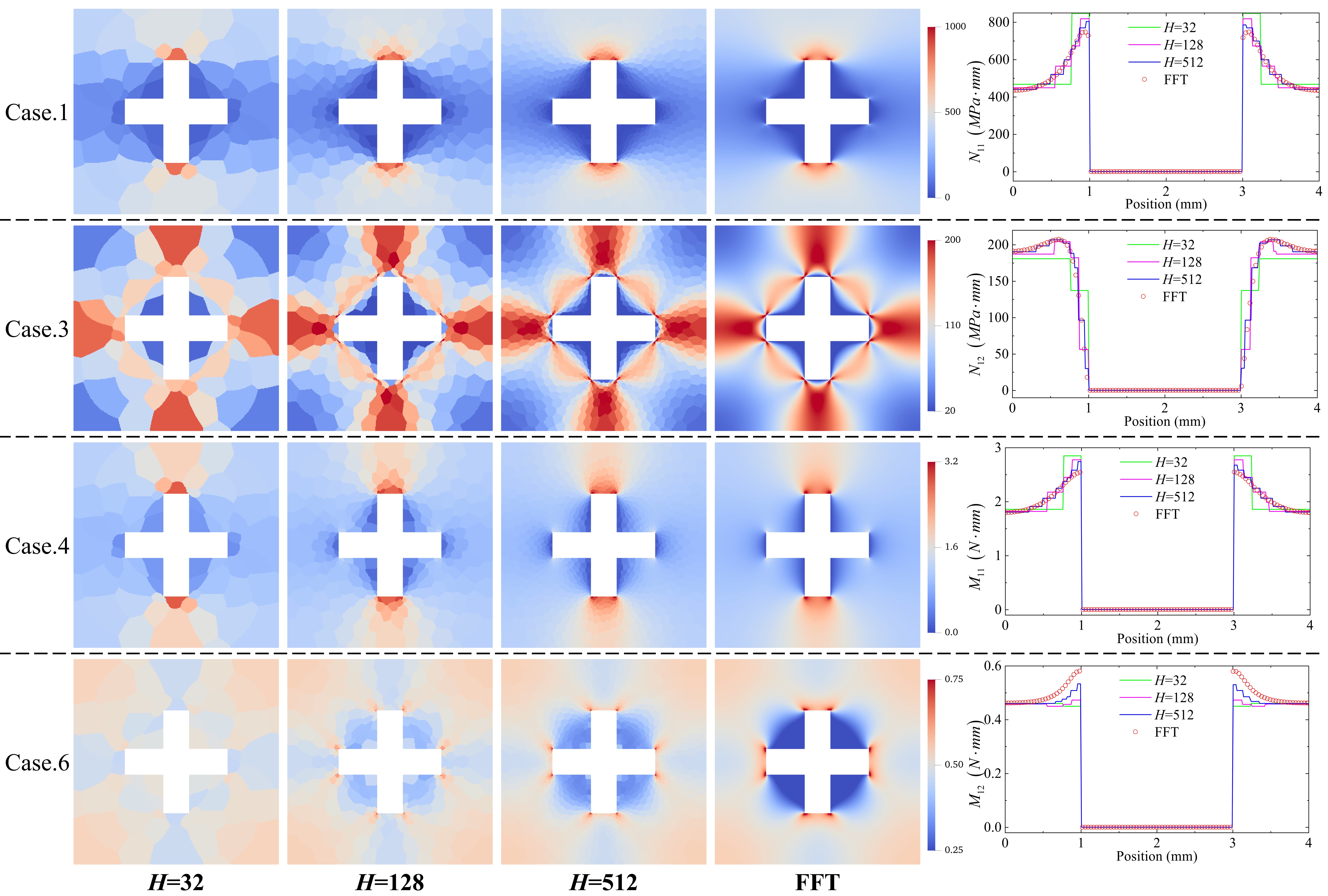}
\caption{\em{Comparison between the FFT and SCA results based on CPT for the symmetric plate.}}
\label{fig:F4}
\end{figure}

\begin{figure}[htbp]
\centering
\includegraphics[width=16cm]{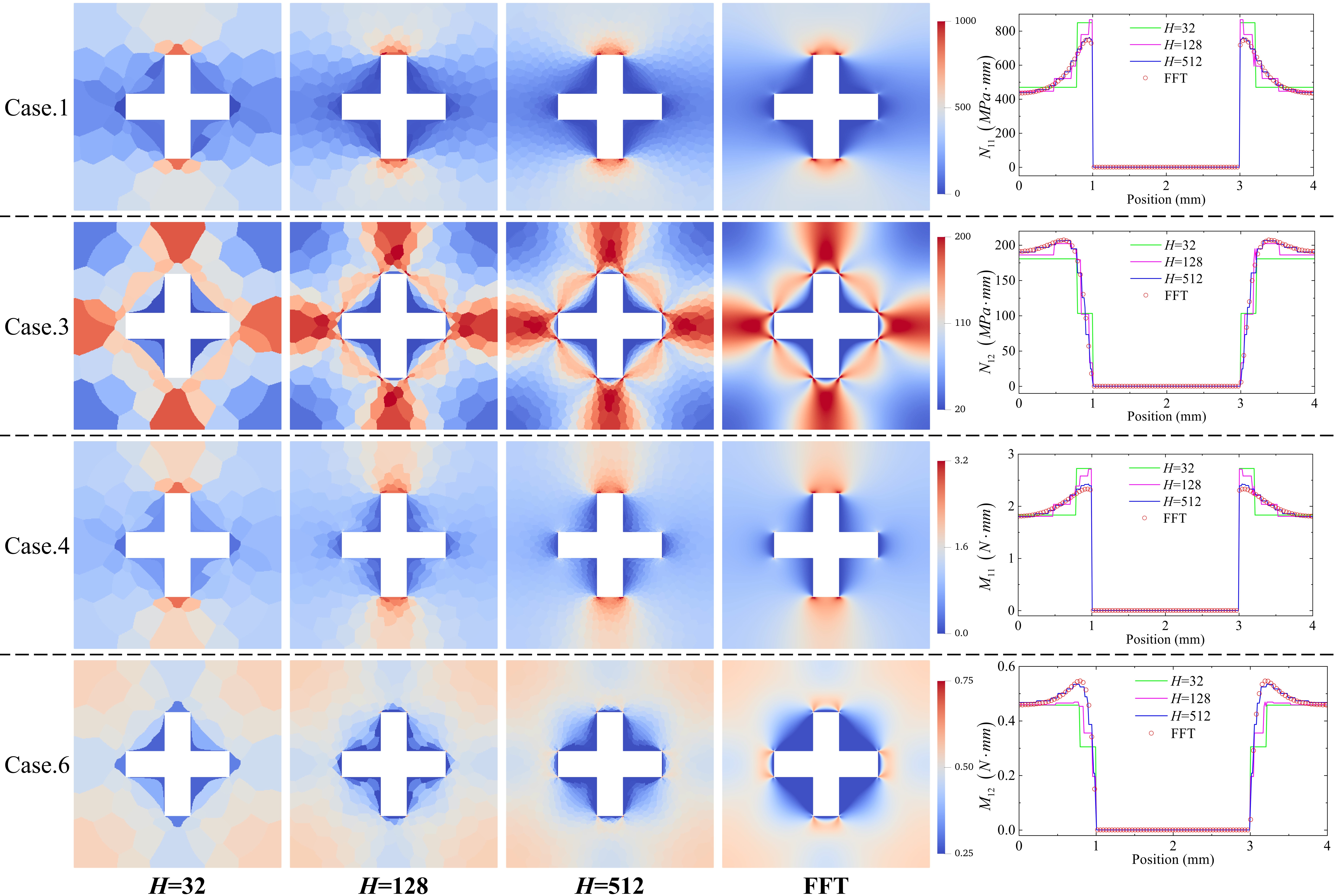}
\caption{\em{Comparison between the FFT and SCA results based on FoPT for the symmetric plate.}}
\label{fig:F5}
\end{figure}

\begin{figure}[htbp]
\centering
\includegraphics[width=16cm]{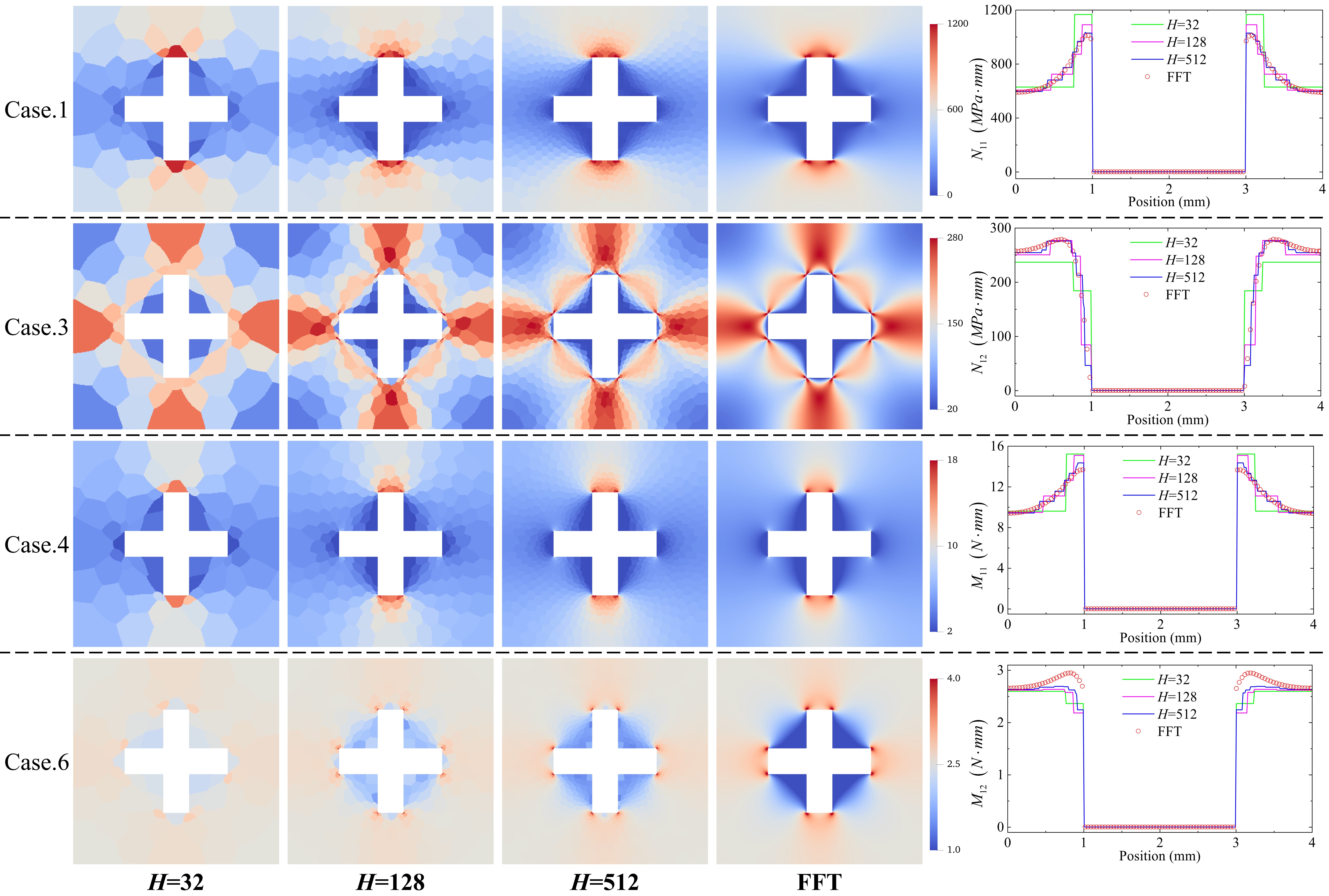}
\caption{\em{Comparison between the FFT and SCA results based on CPT for the asymmetric plate.}}
\label{fig:F6}
\end{figure}

\begin{figure}[htbp]
\centering
\includegraphics[width=16cm]{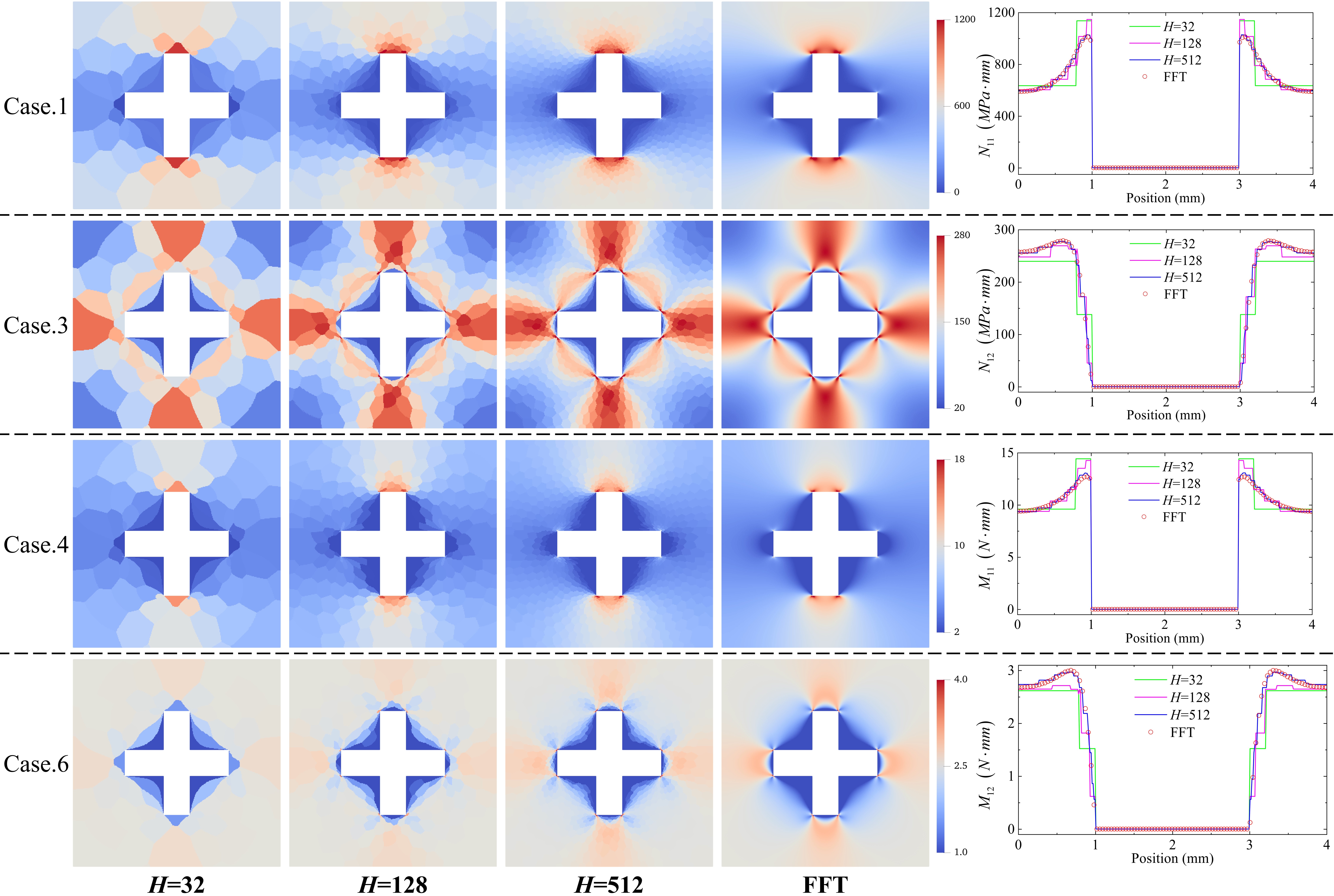}
\caption{\em{Comparison between the FFT and SCA results based on FoPT for the asymmetric plate.}}
\label{fig:F7}
\end{figure}

Figs.~\ref{fig:F4}–\ref{fig:F7} illustrate the stress distributions for four loading cases based on CPT and FoPT, along with the corresponding stress values extracted along the central vertical line at $x = 2\,\mathrm{mm}$. For smaller $H$, the interpolation of the stress field, constrained by the coarse cluster resolution, lacks the accuracy needed to precisely capture stress concentration regions. In this case, the calculated stress range remains within the extreme values obtained from FFT, with SCA overestimating stresses in low-stress regions and underestimating them in high-stress regions. This discrepancy gradually diminishes as $H$ increases, and at $H = 512$, the stress distributions exhibit close agreement. However, the extracted stress values indicate that SCA produces piecewise-constant stress distributions, reflecting its inherent cluster-wise uniform stress assignment. Furthermore, while the distributions of $N$ remain largely consistent between CPT and FoPT, notable differences are observed in $M$. Compared to CPT, FoPT demonstrates superior accuracy in capturing $M$ in SCA calculations, as it incorporates more detailed bending features when constructing the offline clustering database.

\subsection{Linear analysis of twill woven composites}\label{sec:3.2}

\begin{figure}[htbp]
\centering
\includegraphics[width=16cm]{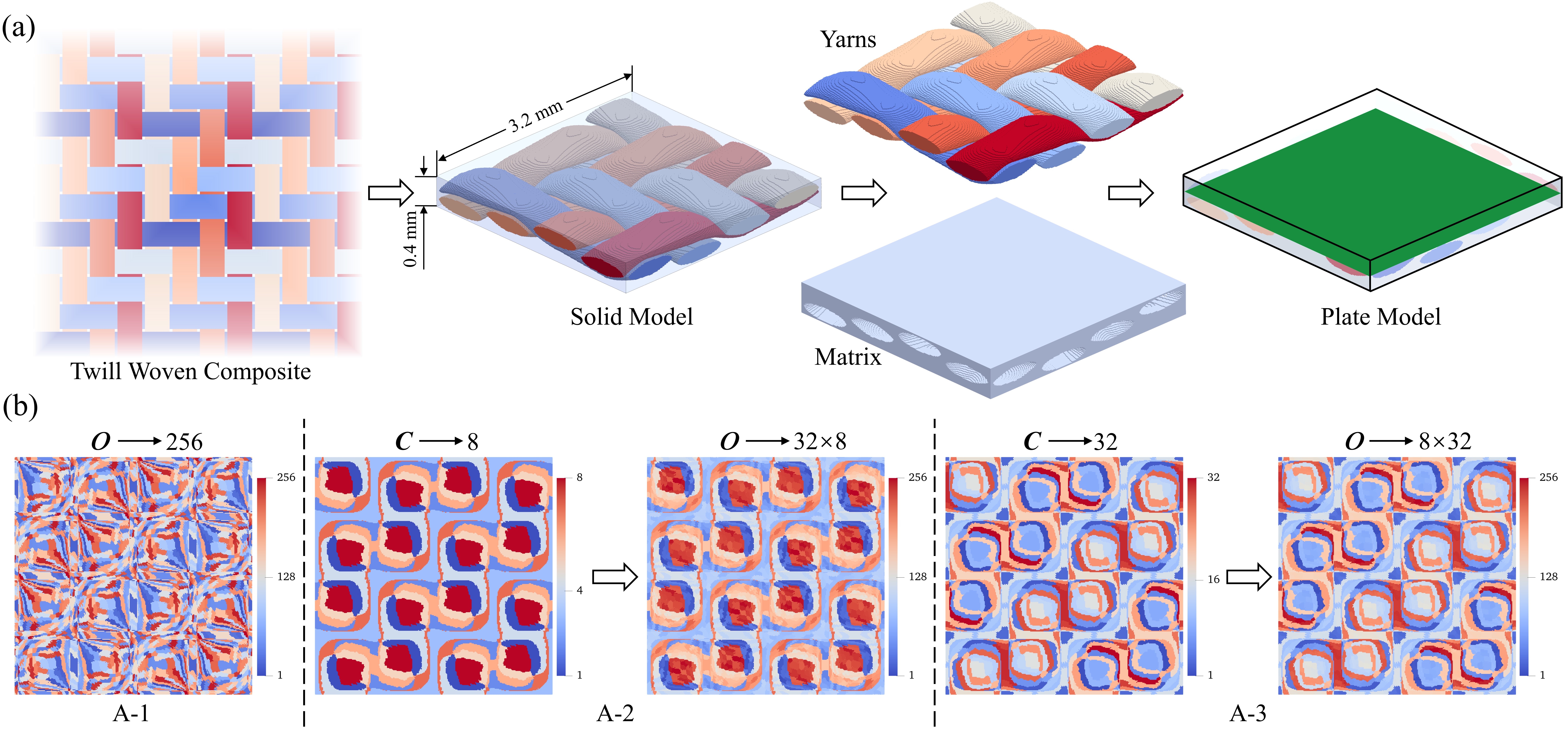}
\caption{\em{Twill weave composite material. (a) plate and solid formulation; (b) three progressive clustering arrangements.}}
\label{fig:F8}
\end{figure}

A twill weave composite material is investigated in this numerical example, as shown in Fig.~\ref{fig:F8}(a), where the macroscopic configuration and the microscopic model are presented. The computational domain is discretised as $320 \times 320 \times 40$. The elastic properties of the yarn and matrix are listed in Table~\ref{tab:T3}. Notably, due to the strongly anisotropic characteristics of woven composites, clustering analysis based solely on the strain concentration tensor may result in excessive variations in mechanical properties within each cluster, potentially compromising computational accuracy. To address this issue, a two-step clustering strategy that incorporates both the material concentration tensor and the strain concentration tensor is commonly employed. Specifically, the initial clustering is performed based on the material concentration tensor $\bm{C}$, followed by secondary clustering of the strain concentration tensor $\bm{O}$ within each sub-cluster. However, existing works have not explicitly clarified the relationship between the clustering configurations of the material concentration tensor and the strain concentration tensor. Based on this context, three progressive clustering arrangements (A-1, A-2, and A-3) are designed for this example, as shown in Fig.~\ref{fig:F8}(b). A-1 directly uses the strain concentration tensor for one-step clustering; A-2 fixes the clustering number of the material concentration tensor at 8 while progressively increasing the sub-cluster number of the strain concentration tensor; conversely, A-3 fixes the sub-cluster number of the strain concentration tensor at 8 while gradually increasing the clustering number of the material concentration tensor. By comparing the computational accuracy of the different schemes (detailed data are provided in Appendix~\ref{sec:A.4}), A-2 is finally adopted as the clustering approach for this example.

\begin{table}[h]
    \renewcommand*{\arraystretch}{1.5}
    \centering
    \caption{Elastic properties of yarn and matrix material \cite{Ref30, Ref49}.}
    \label{tab:T3}      
    \resizebox{0.6\textwidth}{!}{ 
    \begin{tabular}{l c c c c c}
        \hline
        Material & $E_{11}$ (GPa) & $E_{22}$ (GPa) & $\nu_{12}$ & $\nu_{23}$ & $G_{12}$ (GPa) \\
        \hline
        Yarn    & 53.120 & 14.660 & 0.266 & 0.268 & 4.240 \\
        Matrix  & 3.170  & -      & 0.350 & -     & -     \\
        \hline
    \end{tabular}
    }
\end{table}

The homogenised properties of the twill weave composite material are presented in Table~\ref{tab:T4}, based on FFT and SCA with five different cluster numbers $H$. Since the overall structure of the twill weave composite is symmetric with respect to the mid-plane at the macroscopic level, the components $B_{ij} = 0$. The results indicate a significant discrepancy between the homogenised properties calculated using CPT and FoPT, mainly due to the relatively large thickness-to-length ratio of the model ($0.125 > 0.1$). Furthermore, similar to the periodically perforated plate model, as $H$ increases, the homogenised properties gradually converge to those obtained via FFT. For $H = 128$, the error remains within 1\%.

\begin{table}[h]
    \renewcommand*{\arraystretch}{1.5}
    \centering
    \caption{Homogenized properties of the twill weave composite material ($A_{ij}$ [MPa$\cdot$m], $D_{ij}$ [N$\cdot$mm]).}
    \label{tab:T4}      
    \resizebox{\textwidth}{!}{ 
    \begin{tabular}{l c c c c c c c c c c c c}
        \hline
        & \multicolumn{2}{c}{$H=32$} & \multicolumn{2}{c}{$H=64$} & \multicolumn{2}{c}{$H=128$} & \multicolumn{2}{c}{$H=256$} & \multicolumn{2}{c}{$H=512$} & \multicolumn{2}{c}{$\text{FFT}$} \\
        & CPT & FoPT & CPT & FoPT & CPT & FoPT & CPT & FoPT & CPT & FoPT & CPT & FoPT \\
        \hline
        $A_{11}$ & 6.411 & 6.237 & 6.367 & 6.182 & 6.327 & 6.160 & 6.311 & 6.155 & 6.298 & 6.142 & 6.288 & 6.147 \\
        $A_{12}$ & 1.090 & 1.018 & 1.072 & 0.999 & 1.049 & 0.989 & 1.041 & 0.987 & 1.034 & 0.978 & 1.030 & 0.985 \\
        $A_{33}$ & 1.189 & 1.177 & 1.186 & 1.172 & 1.179 & 1.165 & 1.175 & 1.161 & 1.174 & 1.158 & 1.172 & 1.157 \\
        \hline
        $D_{11}$ & 58.162 & 57.146 & 57.361 & 56.191 & 56.636 & 55.799 & 56.333 & 55.722 & 56.113 & 55.331 & 55.898 & 55.078 \\
        $D_{12}$ & 8.417  & 8.792  & 8.346  & 8.640  & 8.405  & 8.613  & 8.438  & 8.603  & 8.441  & 8.542  & 8.430  & 8.517  \\
        $D_{33}$ & 12.481 & 12.096 & 12.469 & 12.050 & 12.468 & 12.034 & 12.464 & 12.025 & 12.461 & 12.021 & 12.456 & 12.048 \\
        \hline
    \end{tabular}
    }
\end{table}

\begin{figure}[htbp]
\centering
\includegraphics[width=16cm]{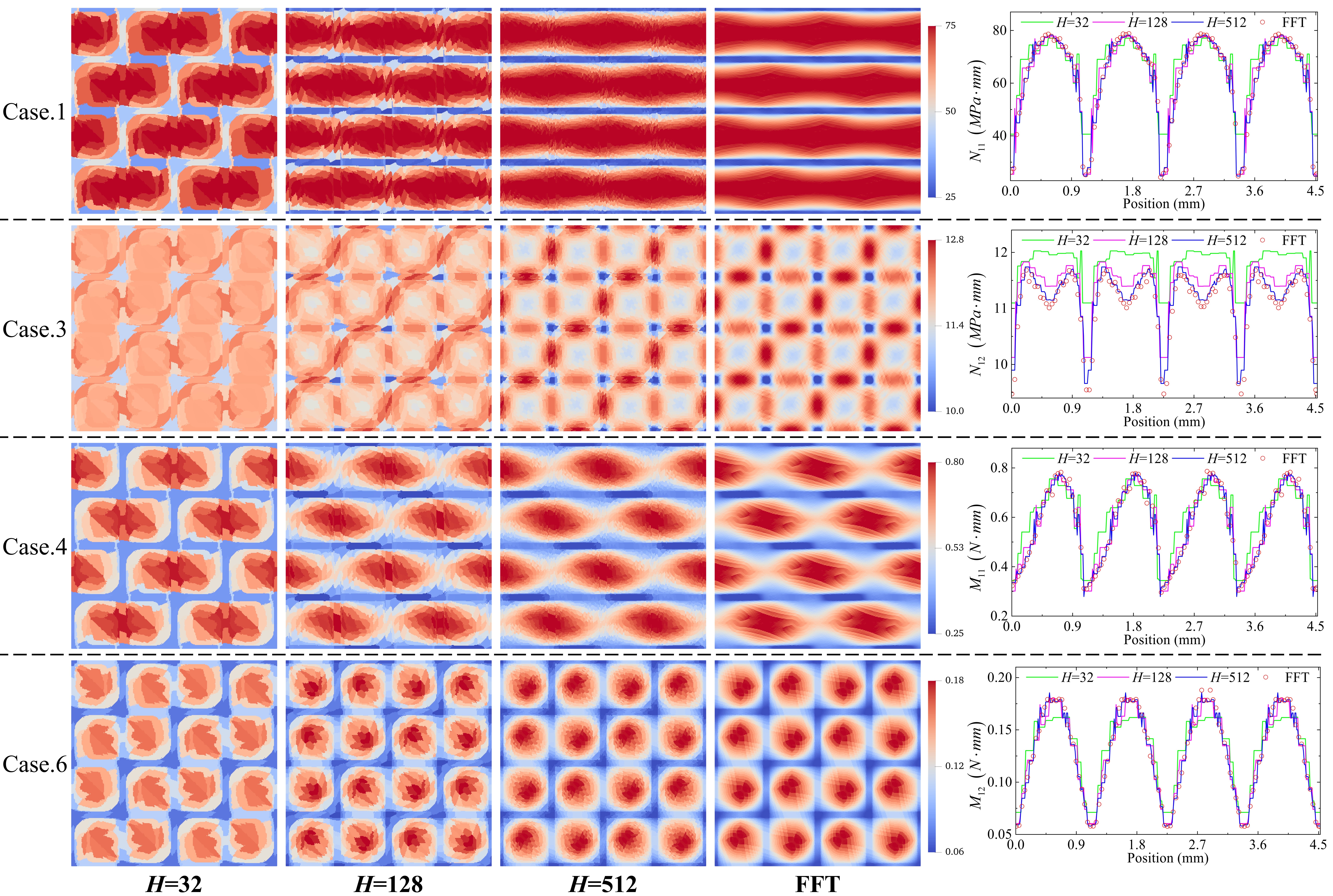}
\caption{\em{Comparison between the FFT and SCA results based on CPT for the twill weave composite material.}}
\label{fig:F9}
\end{figure}

\begin{figure}[htbp]
\centering
\includegraphics[width=16cm]{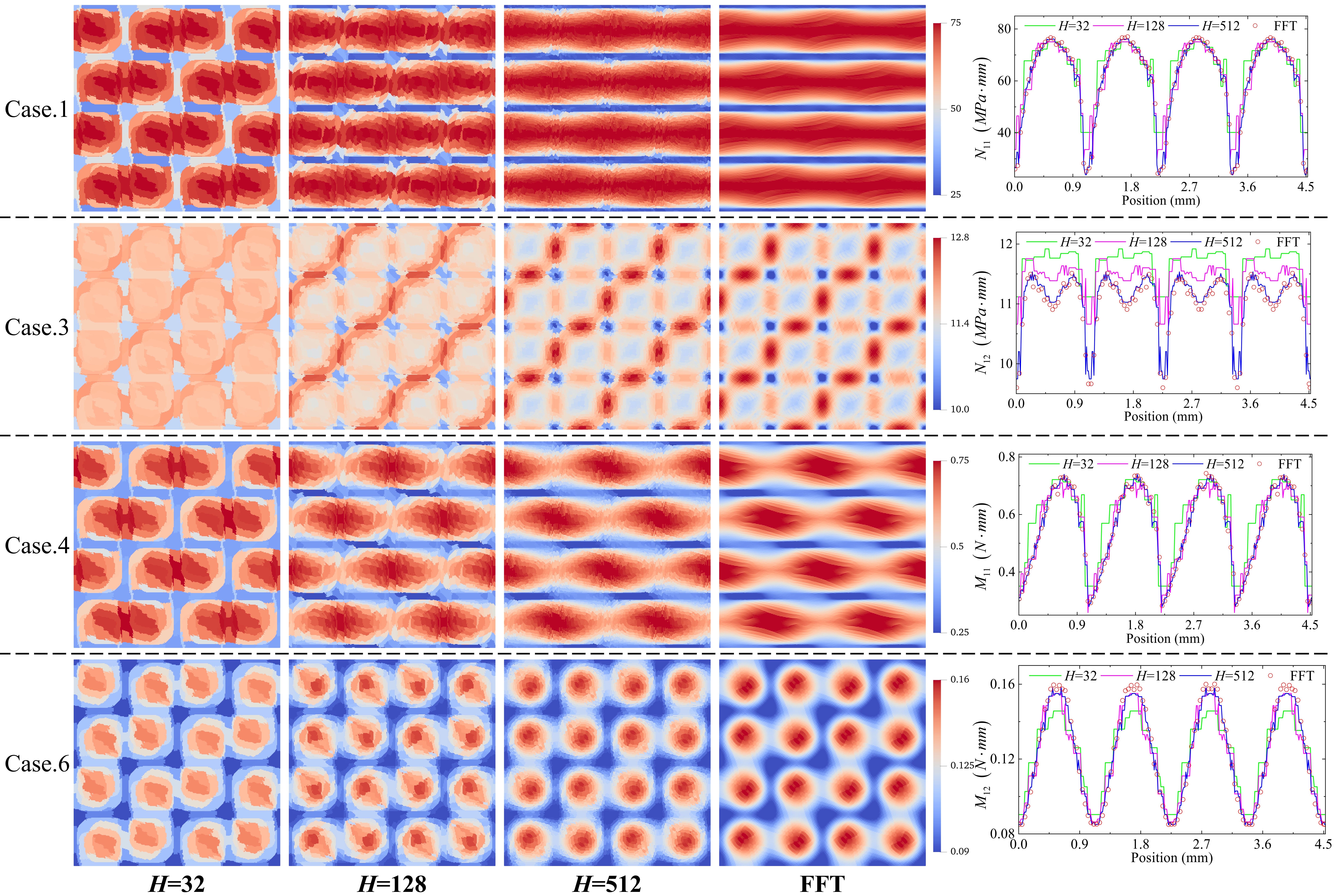}
\caption{\em{Comparison between the FFT and SCA results based on FoPT for the twill weave composite material.}}
\label{fig:F10}
\end{figure}

Fig.~\ref{fig:F9} and Fig.~\ref{fig:F10} present the stress distributions for four loading cases, along with the corresponding stress values extracted along the diagonal line at $x = y$, based on CPT and FoPT, respectively. The results indicate that even for $H = 32$, the SCA method is able to capture the overall stress distribution, as further supported by the extracted stress values along the diagonal. In addition, the agreement between the SCA and FFT results improves as $H$ increases. However, even at $H = 512$, significant local differences remain between the stress field computed by SCA and that obtained using the FFT method, as illustrated by the stress distributions in Case 1. This difference arises from the fact that clustering in SCA is performed based on the strain concentration tensors under six orthogonal load cases, aiming to generalise to arbitrary loading conditions. Consequently, the material points within each cluster do not exhibit strict similarity under a given load case. A rigorous theoretical discussion on the convergence of cluster number in SCA is available in \cite{Ref29}. Furthermore, the extracted stress values show that as $H$ increases, the SCA results gradually converge to those of the FFT method. Regardless of whether CPT or FoPT is used, the stress values extracted from SCA at $H = 128$ are largely consistent with those of FFT, and at $H = 512$ they are nearly identical.

\subsection{Non-linear elasto-plastic analysis of periodic perforated plate}\label{sec:3.3}

In the linear elastic analysis of fixed models, the performance advantages of the SCA method are not fully demonstrated due to its reliance on an "offline–online" two-stage computational framework. Its primary advantage lies in using a single offline calculation to enhance computational efficiency during the online stage. This benefit is particularly important in scenarios requiring repeated computations, such as multi-incremental and multiscale analyses. However, as discussed in Sec.~\ref{sec:2.3.4}, when SCA is used to simulate non-linear behaviour, its prediction accuracy may be affected. Therefore, it is necessary to evaluate the predictive capability of SCA in non-linear simulations. To this end, we present the elasto-plastic (weakly non-linear) analysis of the symmetric perforated plate described in Sec.~\ref{sec:3.1}, under different loading conditions. Since the thickness-to-length ratio of the perforated plate is $0.05 < 0.1$, the following analysis is based solely on the CPT. In this analysis, the elastic stiffness properties remain the same as in Sec.~\ref{sec:3.1}, while the plastic behaviour follows the von Mises yield criterion with the following linear hardening law \cite{Ref47}:

\begin{equation} \label{eq:35}
{{\sigma }_{y}}=0.4+10\cdot p\text{ }\left[ \text{GPa} \right]
\end{equation}
where $\sigma_y$ and $p$ denote the yield stress and equivalent plastic strain, respectively. 

\begin{figure}[htbp]
\centering
\includegraphics[width=12cm]{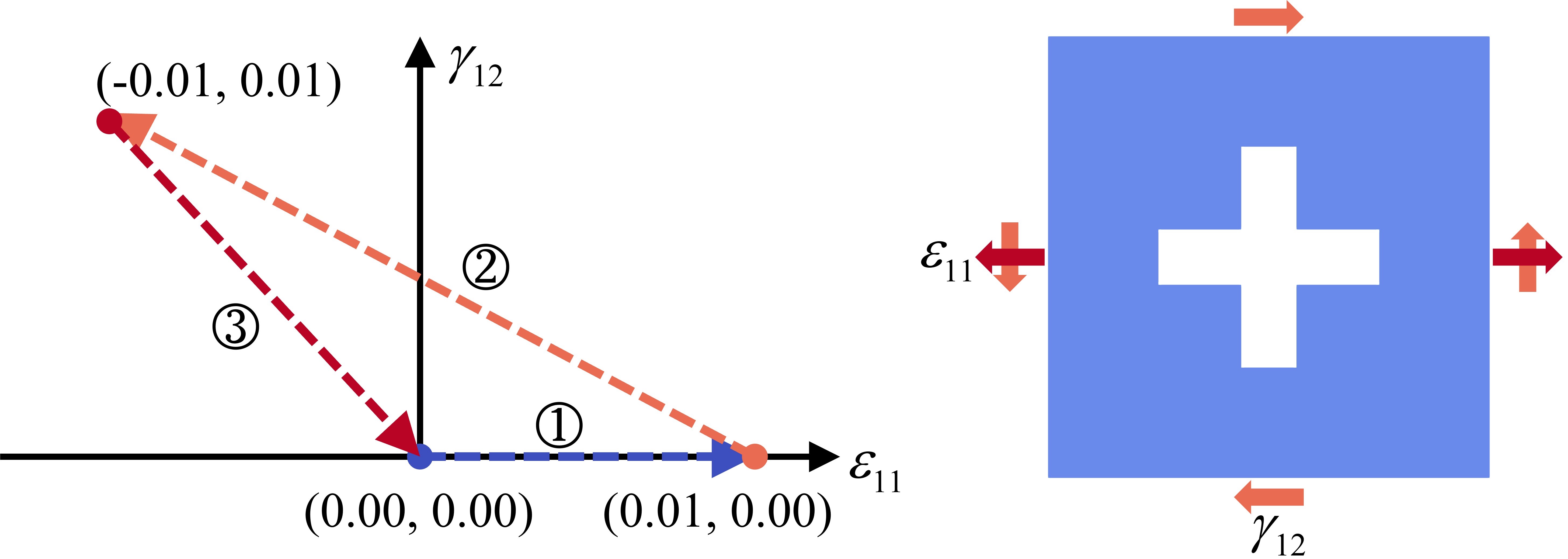}
\caption{\em{Illustrations of the complex loading path.}}
\label{fig:F11}
\end{figure}

Three loading conditions are applied in this example: uniaxial tension, pure shear ($\gamma_{12} = 0.01$), and a complex loading path (as shown in Fig.~\ref{fig:F11}). The first two loading conditions are applied incrementally with 100 uniformly distributed load steps, while the complex loading path is discretised into 400 increments to ensure an accurate representation of the loading process. The boundary conditions for the uniaxial tension case are not explicitly defined by a homogeneous strain loading. Therefore, an \textit{a posteriori} correction procedure is employed to ensure that the following boundary conditions are satisfied:

\begin{equation} \label{eq:36}
\bar{\bm \varepsilon}=\left[ \begin{matrix}   0.01 & \ast \\
   \ast & \ast \end{matrix} \right],\bar{\bm \phi}=\left[ \begin{matrix}   \ast & \ast \\
   \ast & \ast \end{matrix} \right],\bar{\bm N}=\left[ \begin{matrix}   \ast & 0 \\
   0 & 0 \end{matrix} \right],\bar{\bm M}=\left[ \begin{matrix}   0 & 0 \\
   0 & 0 \end{matrix} \right]
\end{equation}

\begin{figure}[htbp]
\centering
\includegraphics[width=14cm]{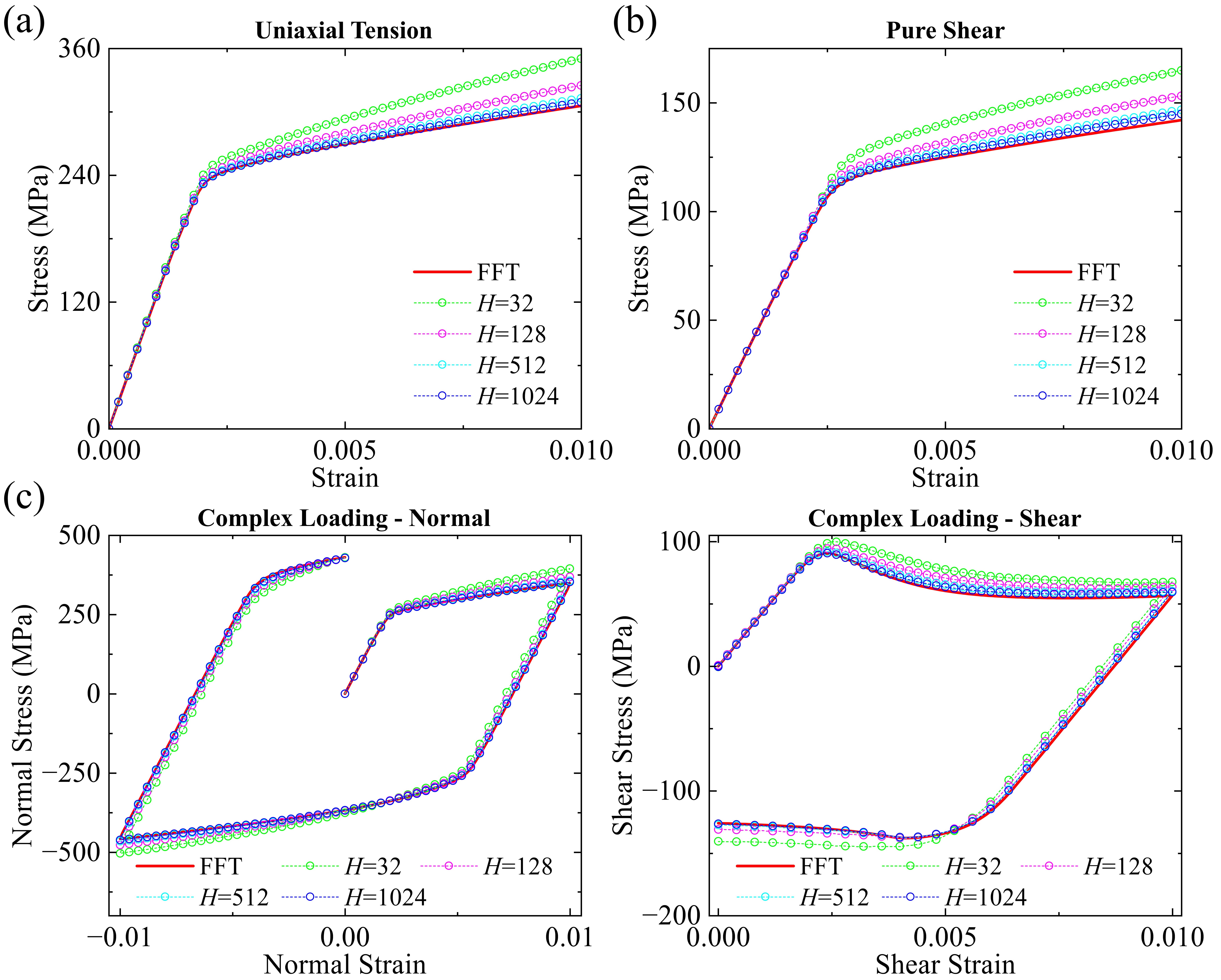}
\caption{\em{Stress-strain curves for elasto-plastic symmetrical perforated plate. (a) uniaxial tension; (b) pure shear; (c) complex loading path.}}
\label{fig:F12}
\end{figure}

Fig.~\ref{fig:F12}(a) and Fig.~\ref{fig:F12}(b) show the stress–strain curves obtained from SCA and FFT calculations under uniaxial tension and pure shear loading, respectively. The results indicate that while the SCA method effectively captures the non-linear elasto-plastic behaviour, its accuracy decreases when $H$ is relatively small. To further illustrate this trend, we include the case with $H = 1024$ and show that increasing the number of clusters consistently improves computational accuracy. Furthermore, Fig.~\ref{fig:F12}(c) shows the normal and tangential stress–strain responses under a complex loading path, where again a high level of agreement between the two methods is observed.

\begin{figure}[htbp]
\centering
\includegraphics[width=16cm]{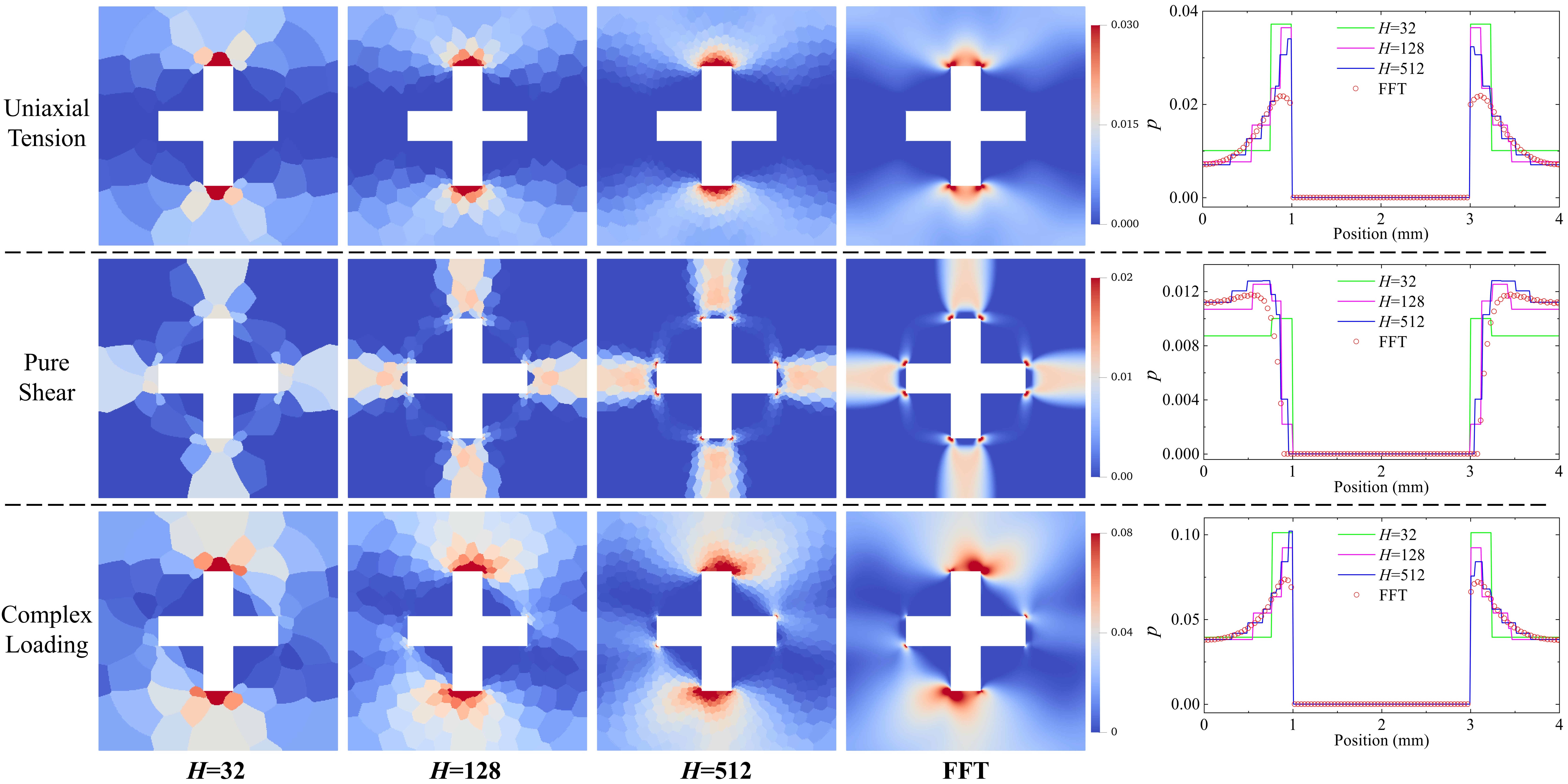}
\caption{\em{Equivalent plastic strain fields for elasto-plastic symmetrical perforated plate.}}
\label{fig:F13}
\end{figure}

Fig.~\ref{fig:F13} shows the final equivalent plastic strain distributions obtained using the SCA and FFT methods under three different loading conditions, along with the corresponding equivalent plastic strain values extracted along the central vertical line. Similar to the linear elastic analysis, SCA is also capable of predicting the equivalent plastic strain field. However, the local distribution appears more diffuse, resulting in a lower peak equivalent plastic strain compared to the FFT results. This is due to the fundamental design principle of SCA, which aims to achieve accurate prediction of globally homogenised mechanical responses while minimising computational information. Within the coarse clustering framework of SCA, the equivalent plastic strain in each cluster is treated as the average of the equivalent plastic strains of all pixels within that cluster. Consequently, high equivalent plastic strain pixels near the pore are averaged with low equivalent plastic strain pixels farther from the pore, thereby mitigating the local strain concentration effect. As a result, the SCA method provides a more conservative estimate, leading to a predicted stress–strain curve that is higher than that obtained using the FFT method. To increase the resolution in local regions, global accuracy can be improved by increasing the total number of clusters. Alternatively, a more efficient approach is to apply a non-linear K-means clustering strategy, where a larger number of clusters are dynamically assigned to critical regions with high strain gradients, such as areas around the pore.

\subsection{Non-linear progressive failure analysis of a practical plain woven composite}\label{sec:3.4}

\begin{figure}[htbp]
\centering
\includegraphics[width=10cm]{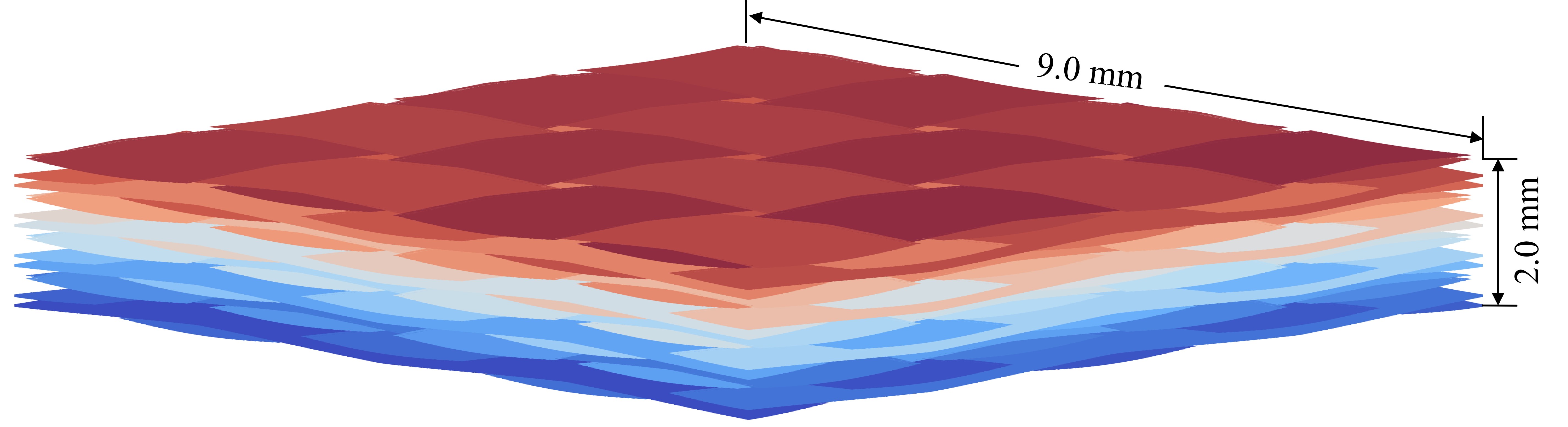}
\caption{\em{Reconstructed RVE model based on the PEM.}}
\label{fig:F14}
\end{figure}

To demonstrate the potential of the proposed method for solving non-linear problems and handling complex structures, a progressive failure analysis of a multilayer plain woven composite is presented in this section. In addition, experimental validation is performed to further verify the effectiveness of the proposed method. The numerical model of the plain woven composite is constructed based on real material data obtained from microscopic images. A modelling strategy known as the projected element method (PEM) is used to reconstruct the RVE of the composite, accurately capturing the deformation of the yarns introduced during the manufacturing process. The reconstructed composite structure and corresponding RVE, including the plate formulation, are shown in Fig.~\ref{fig:F14}.

The non-linear behaviour of the composite is simulated using the damage models described in \cite{Ref36}. The resin phase follows a rate-dependent damage model, while the yarn phase follows the Hashin damage model as the failure initiation criterion. The post-damage behaviour of the composite is described by a Weibull distribution-based model; the specific material parameters can be found in Appendix~\ref{sec:A.5}. The in-plane discretisation resolution of the model is set to $128 \times 128$. Since the thickness-to-length ratio of the model is $0.222 > 0.1$, only the FoPT is used for the simulations. Uniaxial tension and bending loads are applied, both with a loading rate of $10^{-4}\ \mathrm{s}^{-1}$.

\begin{figure}[htbp]
\centering
\includegraphics[width=0.99\textwidth]{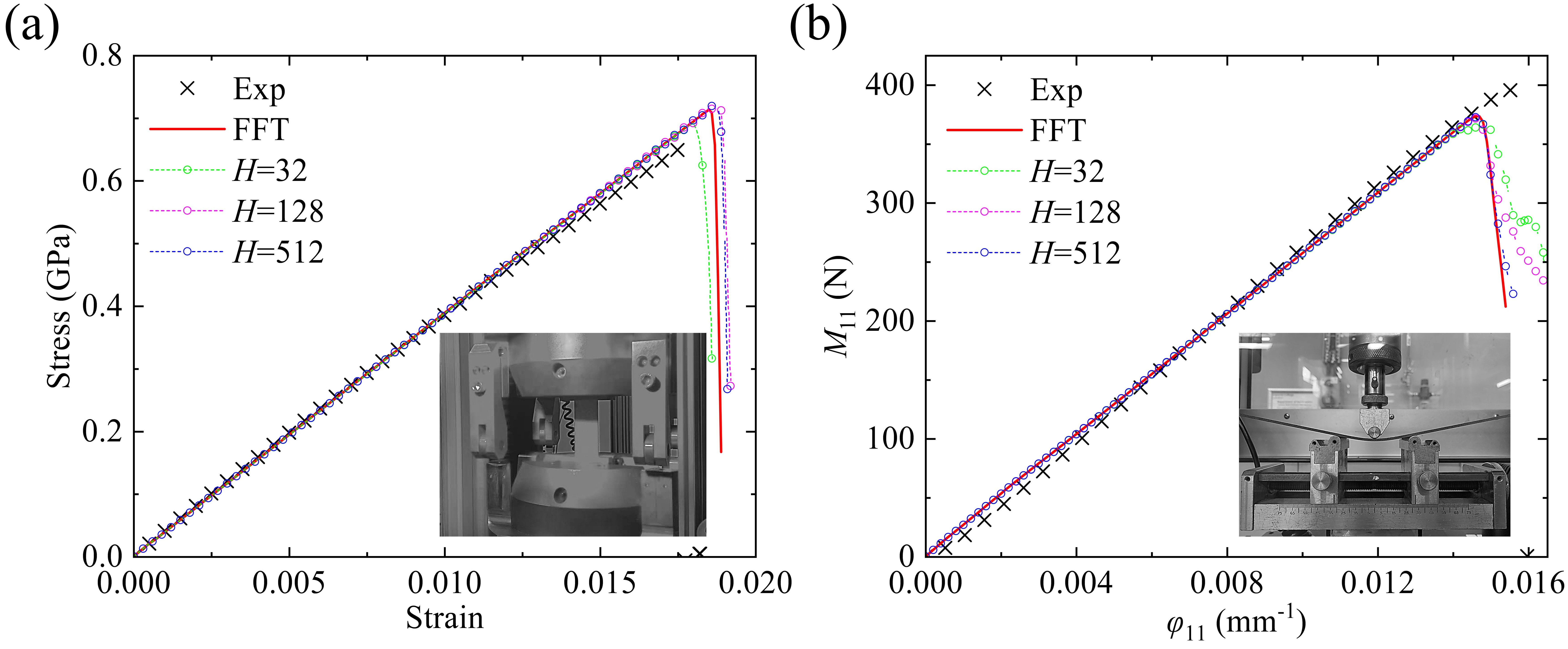}
\caption{\em{Stress-strain curves under different loading conditions: (a) Tensile loading; (b) Bending loading.}}
\label{fig:F15}
\end{figure}

Fig.~\ref{fig:F15} shows the stress–strain curves obtained using the SCA and FFT methods under the two different loading conditions, along with the experimental results. The results show consistency across all methods, with the calculated modulus closely matching the experimental data. Under tensile loading, the calculated tensile strength is approximately 8\% higher than the experimental average, while under bending loading, the calculated strength is approximately 5\% lower than the experimental average.

\section{Discussion} \label{sec:Discussion}

\subsection{Sensitivity analysis of $G^0$}\label{sec:4.1}

The accuracy of SCA strongly depends on the self-consistent update of the reference medium. In this study, the projection-based scheme described in Appendix~\ref{sec:A.3} is used to update the reference medium. However, for the FoPT problem, as formulated in Eq.~(\ref{eq:24}), the material parameters ($\mu_b^0$ and $G^0$) of the reference medium in the Green's operator are coupled to the discrete frequencies. This coupling requires the Green's operator—and therefore the interaction tensor—to be recalculated at each update step. As discussed in Sec.~\ref{sec:2.3.2}, if $\mu_b^0$ and $G^0$ satisfy the assumption in Eq.~(\ref{eq:26}), the coupled term ($\rho$) can be treated as a coefficient, eliminating the need to recalculate the interaction tensor. 

Existing sensitivity analyses in the context of the FFT method indicate that $\mu_b^0$ exhibits significant sensitivity, and the term $2\mu_b^0 + \lambda_b^0$ influences the overall convergence behaviour of the computation. Therefore, when solving the FoPT problem with SCA, we employ a self-consistent update strategy for $\mu_b^0$ to ensure its accuracy, while $G^0$ is updated accordingly based on the condition imposed by Eq.~(\ref{eq:26}). Although $G^0$ has been shown to be relatively insensitive in the FFT method, its sensitivity in SCA remains unclear. The following is the selection strategy for $G^0$ in the FFT method:
\begin{equation} \label{eq:37}
{{G}^{0}}=\beta(G^0) \left( \underset{x\in \Omega }{\mathop{\max }}\,\text{eig}\left[\bm S\left(\bm x \right) \right]+\underset{x\in \Omega }{\mathop{\min }}\,\text{eig}\left[\bm S\left( \bm x \right) \right] \right)
\end{equation}

Here, $\text{eig}(\cdot)$ denotes the eigenvalues of a matrix, and $\beta(G^0)$ represents a coefficient associated with the selection of the reference medium. By adjusting this coefficient, $G^0$ can be modified. To comprehensively investigate the sensitivity of $G^0$, the twill weave composite material introduced in Section~3.2 is employed. Several values of $\beta(G^0)$ are selected, ranging from $10^{-2}$ to $10^{2}$, while all other parameters of the reference medium are kept fixed.

\begin{figure}[htbp]
\centering
\includegraphics[width=15cm]{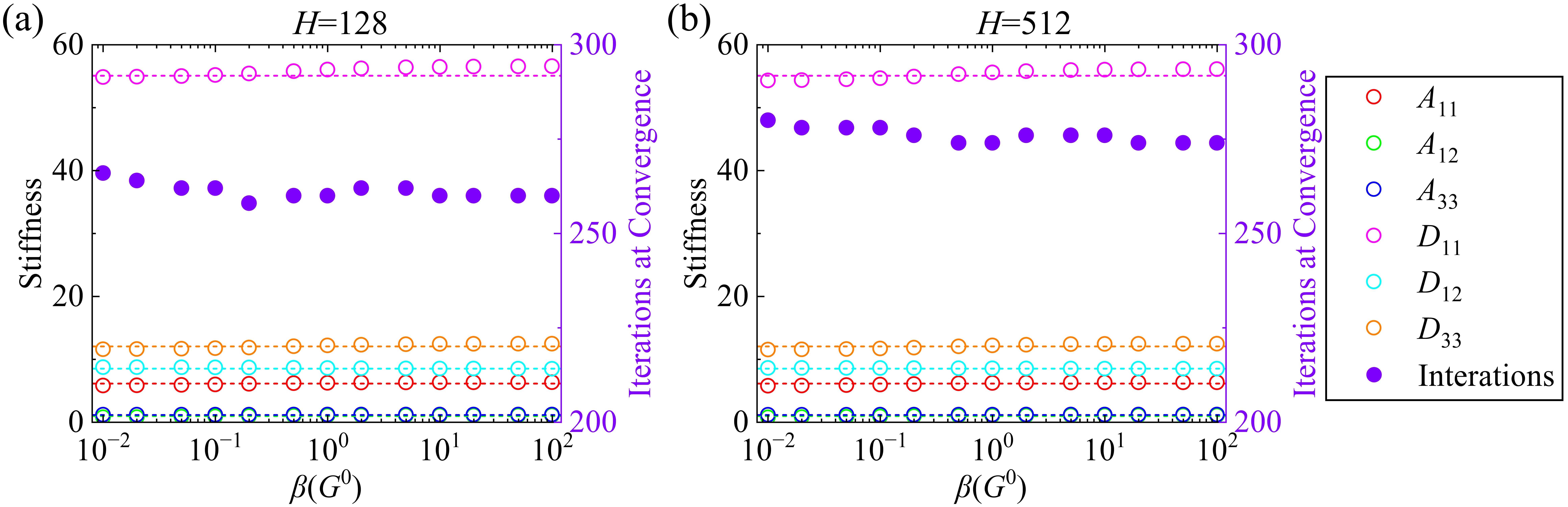}
\caption{\em{Homogenized properties and iterations for the different $G^0$. (a) $H=128$; (b) $H=512$.}}
\label{fig:F16}
\end{figure}

Fig.~\ref{fig:F16} illustrates the homogenised properties obtained from the SCA calculations for two different cluster numbers ($H = 128$ and $H = 512$), along with the total number of iterations required under six orthogonal loading conditions. The dashed lines in the figure represent the reference homogenised properties computed using the FFT method. All stiffness values are in close agreement with the reference values, indicating that variations in $G^0$ do not affect the computational accuracy. In addition, the total number of Newton–Raphson iterations required to solve the six load cases remains nearly constant for different $G^0$. This conclusion holds for different clustering numbers, demonstrating that the computation of membrane and bending responses is insensitive to $G^0$. Consequently, the assumption used in Eq.~(\ref{eq:26}) to calculate the FoPT is valid.

\subsection{Computation time}\label{sec:4.2}

The computational efficiency of the developed SCA, as a reduced-order model (ROM) for heterogeneous plates, is systematically evaluated through comparative studies with direct numerical simulation (DNS) using the FFT-based homogenisation method. All numerical tests were conducted on a workstation equipped with an Intel(R) Core(TM) i7-11700 CPU. 

Table~\ref{tab:T5} summarises the cumulative computational time for six loading cases across three examples: Example 1 (symmetrical perforated plate in Sec.~\ref{sec:3.1}), Example 2 (asymmetrical perforated plate in Sec.~\ref{sec:3.1}), and Example 3 (twill-woven composite in Sec.~\ref{sec:3.2}). SCA demonstrates superior computational efficiency, achieving over 10× speed-up compared to FFT even at $H = 512$. The speed-up exceeds two orders of magnitude when $H = 128$ is used, with acceptable solution accuracy (error $<1\%$). Similar performance advantages are observed for elasto-plastic analyses of perforated plates under three loading conditions (Table~\ref{tab:T6}), where SCA achieves more than twice the computational efficiency relative to FFT at $H = 1024$. Notably, the computational time exhibits quadratic scaling with respect to the cluster number $H$ ($T \propto H^2$), consistent with numerical observations in classical SCA frameworks \cite{Ref30}. These results highlight SCA's exceptional ability to balance computational accuracy and efficiency—especially considering that the FFT-based DNS method itself is widely recognised as a computationally efficient homogenisation technique within the computational mechanics community. The speed-up achieved stems from the intrinsic dimensionality reduction of SCA’s data-driven architecture, which bypasses intensive local field computations through effective clustering-based homogenisation.

\begin{table}[h]
    \renewcommand*{\arraystretch}{1.5}
    \centering
    \caption{Computation time of Example 1–Example 3 (Unit: [s]).}
    \label{tab:T5}      
    \resizebox{\textwidth}{!}{ 
    \begin{tabular}{l c c c c c c c c c c c c}
        \hline
        & \multicolumn{2}{c}{$H=32$} & \multicolumn{2}{c}{$H=64$} & \multicolumn{2}{c}{$H=128$} & \multicolumn{2}{c}{$H=256$} & \multicolumn{2}{c}{$H=512$} & \multicolumn{2}{c}{$\text{FFT}$} \\
        & CPT & FoPT & CPT & FoPT & CPT & FoPT & CPT & FoPT & CPT & FoPT & CPT & FoPT \\
        \hline
        Example 1& 0.002 & 0.004 & 0.007 & 0.011 & 0.055 & 0.083 & 0.231 & 0.341 & 1.249 & 1.675 & 14.304 & 38.781 \\
        Example 2& 0.017 & 0.028 & 0.055 & 0.077 & 0.386 & 0.565 & 1.688 & 2.300 & 8.730 & 12.432 & 24.186 & 65.460 \\
        Example 3& 0.006 & 0.009 & 0.021 & 0.030 & 0.134 & 0.213 & 0.543 & 0.888 & 2.696 & 3.980 & 20.170 & 30.880 \\
        \hline
    \end{tabular}
    }
\end{table}

\begin{table}[h]
    \renewcommand*{\arraystretch}{1.5}
    \centering
    \caption{Computation time of elasto-plastic symmetrical perforated plate (Unit: [s]).}
    \label{tab:T6}      
    \resizebox{\textwidth}{!}{ 
    \begin{tabular}{l c c c c c c c}
        \hline
        & $H=32$ & $H=64$ & $H=128$ & $H=256$ & $H=512$ & $H=1024$ & $\text{FFT}$\\
        \hline
        Uniaxial Tension & 0.044 & 0.111 & 0.407 & 1.665 & 5.144 & 28.972 & 70.148 \\
        Pure Shear & 0.035 & 0.094 & 0.348 & 1.515 & 4.473 & 20.821 & 52.828 \\
        Complex Loading & 0.159 & 0.382 & 1.562 & 6.479 & 18.323 & 112.496 & 352.827 \\
        \hline
    \end{tabular}
    }
\end{table}

The computation time required for DNS generally increases with model resolution. In contrast, the SCA features a computational cost intrinsically governed by the number of clusters, as its degrees of freedom correspond directly to the discretised cluster quantity. To systematically validate this hypothesis, five different discretisations are implemented for the symmetrical perforated plate model in Example 1. Both SCA and FFT methods are employed to compute six loading cases under CPT. The computation time and error are summarised in Table~\ref{tab:T7}, where the error represents the mean absolute error of the homogenised properties relative to the highest-resolution FFT results. The computational results show that FFT computation time increases progressively with higher resolution, while the SCA computation time remains nearly constant across all discretisations. This confirms that the computational cost of SCA is independent of model resolution, enabling refined discretisation to capture intricate geometric features without incurring additional computational cost. Furthermore, the FFT error decreases systematically with increasing resolution, while the SCA error remains essentially unchanged across different discretisations. The SCA error exhibits a decreasing trend with increasing cluster quantity ($H$), which conclusively verifies that its computational degrees of freedom depend solely on the number of clusters rather than the model resolution.

\begin{table}[h]
    \renewcommand*{\arraystretch}{1.5}
    \centering
    \caption{Computation time and error of Example 1 for different discretizations (Error: [\%]), Time: [s]).}
    \label{tab:T7}      
    \resizebox{\textwidth}{!}{ 
    \begin{tabular}{l cc cc cc cc cc cc}
        \hline
        Resolution & \multicolumn{2}{c}{$H=32$} & \multicolumn{2}{c}{$H=64$} & \multicolumn{2}{c}{$H=128$} & \multicolumn{2}{c}{$H=256$} & \multicolumn{2}{c}{$H=512$} & \multicolumn{2}{c}{$\text{FFT}$} \\
        & Err & Time & Err & Time & Err & Time & Err & Time & Err & Time & Err & Time \\
        \hline
        80$\times$80 &  2.410&  0.0027&  1.924&  0.0077&  1.555&  0.0553&  1.133&  0.2319&  0.747&  1.2113&  0.129&  1.3926\\
        160$\times$160 &  2.081&  0.0026&  2.260&  0.0074&  1.252&  0.0548&  1.051&  0.2298&  0.718&  1.2211&  0.087&  2.5978\\
        240$\times$240 &  2.165&  0.0025&  2.228&  0.0082&  1.357&  0.0554&  1.097&  0.2280&  0.804&  1.1940&  0.056&  6.16003\\
        320$\times$320 &  2.138&  0.00249&  2.323&  0.0074&  1.215&  0.0553&  0.999&  0.2307&  0.753&  1.2495&  0.028&  14.3043\\
        400$\times$400 &  2.128&  0.0026&  2.276&  0.0078&  1.180&  0.0547&  1.018&  0.2329&  0.748&  1.1895&  0&  24.9872\\
        \hline
    \end{tabular}
    }
\end{table}

\section{Conclusions} \label{sec:conclusions}

This study successfully develops a self-consistent clustering analysis (SCA) framework that integrates classical plate theory (CPT) and first-order shear deformation plate theory (FoPT), providing an efficient unit cell solution for heterogeneous plate structures. By extending the theoretical framework, the proposed reduced-order model significantly enhances computational efficiency and applicability for heterogeneous plates. Numerical validations demonstrate that the method accurately captures the non-linear behaviour of both isotropic and anisotropic materials, including elastoplastic and progressive failure mechanisms. Compared to direct numerical simulation (DNS), the proposed approach achieves a computational error below 1\% with a limited number of clusters while delivering over a 100-fold speed-up. 

Although the present work focuses on mechanical boundary value problems, the modular design of this framework allows seamless extension to complex multiphysics scenarios such as thermo-mechanical coupling and micropolar continuum mechanics. The findings of this study provide an innovative methodological foundation for various cutting-edge research domains, particularly in:  
(1) multiscale mechanical property prediction of composite materials,  
(2) topology optimisation of porous materials \cite{Ref50},  
(3) intelligent design of mechanical metamaterials, and  
(4) data-driven modelling.  
These directions will serve as key focal points for future research.

\section*{Acknowledgements}

This work was financially supported by the National Key Research and Development Program of China (2022YFB3707800).

\cleardoublepage

\appendix
\section{Appendix}

\subsection{Green's operator of CPT and FoPT} \label{sec:A.1}

The Green's operator of CPT consists of two components, and its explicit expression in Fourier space is as follows:

\begin{equation} \label{eq:A1}
\left\{ \begin{aligned}
  & \hat{G}_{khij}^{0(NN)}=\frac{{{\delta }_{ki}}{{\xi }_{h}}{{\xi }_{j}}+{{\delta }_{hi}}{{\xi }_{k}}{{\xi }_{j}}+{{\delta }_{kj}}{{\xi }_{h}}{{\xi }_{i}}+{{\delta }_{hj}}{{\xi }_{k}}{{\xi }_{i}}}{4\mu _{m}^{0}{{\left| \xi  \right|}^{2}}}-\frac{\lambda _{m}^{0}+\mu _{m}^{0}}{\mu _{m}^{0}\left( \lambda _{m}^{0}+2\mu _{m}^{0} \right)}\frac{{{\xi }_{k}}{{\xi }_{h}}{{\xi }_{i}}{{\xi }_{j}}}{{{\left| \xi  \right|}^{4}}} \\ 
  & \hat{G}_{khij}^{0(MM)}=\frac{{{\xi }_{i}}{{\xi }_{j}}{{\xi }_{k}}{{\xi }_{h}}}{\left( \lambda _{b}^{0}+2\mu _{b}^{0} \right){{\left| \xi  \right|}^{4}}} \\ 
\end{aligned} \right.
\end{equation}

The Green's operator of FoPT consists of five components, and its explicit expression in Fourier space is as follows:

\begin{equation} \label{eq:A2}
\left\{ \begin{aligned}
  & \hat{G}_{khij}^{0(NN)}=\frac{{{\delta }_{ki}}{{\xi }_{h}}{{\xi }_{j}}+{{\delta }_{hi}}{{\xi }_{k}}{{\xi }_{j}}+{{\delta }_{kj}}{{\xi }_{h}}{{\xi }_{i}}+{{\delta }_{hj}}{{\xi }_{k}}{{\xi }_{i}}}{4\mu _{m}^{0}{{\left| \xi  \right|}^{2}}}-\frac{\lambda _{m}^{0}+\mu _{m}^{0}}{\mu _{m}^{0}\left( \lambda _{m}^{0}+2\mu _{m}^{0} \right)}\frac{{{\xi }_{k}}{{\xi }_{h}}{{\xi }_{i}}{{\xi }_{j}}}{{{\left| \xi  \right|}^{4}}} \\ 
  & \hat{G}_{khij}^{0(MM)}=\frac{{{\delta }_{ih}}{{\xi }_{j}}{{\xi }_{k}}+{{\delta }_{ik}}{{\xi }_{h}}{{\xi }_{j}}}{2\left( \mu _{b}^{0}{{\left| \xi  \right|}^{2}}+{{G}^{0}} \right)}-\frac{{{\xi }_{h}}{{\xi }_{i}}{{\xi }_{j}}{{\xi }_{k}}}{{{\left| \xi  \right|}^{2}}}\left[ \frac{1}{\mu _{b}^{0}{{\left| \xi  \right|}^{2}}+{{G}^{0}}}-\frac{1}{\left( \lambda _{b}^{0}+2\mu _{b}^{0} \right){{\left| \xi  \right|}^{2}}} \right] \\ 
  & \hat{G}_{khi}^{0(MQ)}=i\left[ \frac{{{\delta }_{ih}}{{\xi }_{k}}+{{\delta }_{ik}}{{\xi }_{h}}}{2\left( \mu _{b}^{0}{{\left| \xi  \right|}^{2}}+{{G}^{0}} \right)}-\frac{{{\xi }_{h}}{{\xi }_{i}}{{\xi }_{k}}}{\left( \mu _{b}^{0}{{\left| \xi  \right|}^{2}}+{{G}^{0}} \right){{\left| \xi  \right|}^{2}}} \right] \\ 
  & \hat{G}_{kij}^{0(QM)}=i\left[ \frac{{{\xi }_{i}}{{\xi }_{j}}{{\xi }_{k}}}{\left( \mu _{b}^{0}{{\left| \xi  \right|}^{2}}+{{G}^{0}} \right){{\left| \xi  \right|}^{2}}}-\frac{{{\delta }_{ik}}{{\xi }_{j}}}{\mu _{b}^{0}{{\left| \xi  \right|}^{2}}+{{G}^{0}}} \right] \\ 
  & \hat{G}_{ki}^{0(QQ)}=\frac{{{\delta }_{ik}}}{\mu _{b}^{0}{{\left| \xi  \right|}^{2}}+{{G}^{0}}}-\frac{{{\xi }_{i}}{{\xi }_{k}}}{{{\left| \xi  \right|}^{2}}}\left( \frac{1}{\mu _{b}^{0}{{\left| \xi  \right|}^{2}}+{{G}^{0}}}-\frac{1}{{{G}^{0}}} \right) \\ 
\end{aligned} \right.
\end{equation}

Here,  $\lambda _{m}^{0}$, $\mu _{m}^{0}$, $\lambda _{b}^{0}$, $\mu _{b}^{0}$, and ${G}^{0}$ represent the Lame constants and shear modulus of the reference medium. It should be noted that the $\hat{G}_{khij}^{0(NN)}$ terms are identical for both CPT and FoPT, whereas the $\hat{G}_{khij}^{0(MM)}$ terms are different.

\subsection{Cluster-based reduced-order modeling for CPT} \label{sec:A.2}

First, the Lippmann-Schwinger Eq. ( \ref{eq:6} ) is written in an incremental form:

\begin{equation} \label{eq:A3}
\left\{ \begin{aligned}
  & \Delta \boldsymbol{\varepsilon} \left( \boldsymbol{x} \right)+\int_{\Omega }{{{\mathbb{G}}^{0\left( NN \right)}}\left( \boldsymbol{x}-\boldsymbol{y} \right):\left[ \Delta \boldsymbol{N} \left( \boldsymbol{y} \right)-{{\mathbb{A}}^{0}}:\Delta \boldsymbol{\varepsilon} \left( \boldsymbol{y} \right) \right]d{y}}-\Delta \bar{\boldsymbol{\varepsilon}} = 0 \\ 
  & \Delta \boldsymbol{\phi} \left( \boldsymbol{x} \right)+\int_{\Omega }{{{\mathbb{G}}^{0\left( MM \right)}}\left( \boldsymbol{x}-\boldsymbol{y} \right):\left[ \Delta \boldsymbol{M} \left( \boldsymbol{y} \right)-{{\mathbb{D}}^{0}}:\Delta \boldsymbol{\phi} \left( \boldsymbol{y} \right) \right]d{y}}-\Delta \bar{\boldsymbol{\phi}} = 0 \\ 
\end{aligned} \right.
\end{equation}

The RVE model is divided into \textit{H} material clusters. The volume averaging of Eq. ( \ref{eq:A3} ) is performed within the \textit{I}-th cluster as follows:

\begin{equation} \label{eq:A4}
\left\{ \begin{aligned}
  & \frac{1}{{{c}^{I}}\left| \Omega  \right|}\int_{\Omega }{{{{\chi}}^{I}}\left( \boldsymbol{x} \right)\Delta \boldsymbol{\varepsilon} \left( \boldsymbol{x} \right)d{x}} \\ & +\frac{1}{{{c}^{I}}\left| \Omega  \right|}\iint_{\Omega }{{{{\chi}}^{I}}\left( \boldsymbol{x} \right){{\mathbb{G}}^{0\left( NN \right)}}\left( \boldsymbol{x}-\boldsymbol{y} \right):\left[ \Delta \boldsymbol{N} \left( \boldsymbol{y} \right)-{{\mathbb{A}}^{0}}:\Delta \boldsymbol{\varepsilon} \left( \boldsymbol{y} \right) \right]d{y}d{x}}-\Delta \bar{\boldsymbol{\varepsilon}} = 0 \\ 
  & \frac{1}{{{c}^{I}}\left| \Omega  \right|}\int_{\Omega }{{{{\chi}}^{I}}\left( \boldsymbol{x} \right)\Delta \boldsymbol{\phi} \left( \boldsymbol{x} \right)d{x}}  \\&+\frac{1}{{{c}^{I}}\left| \Omega  \right|}\iint_{\Omega }{{{{\chi}}^{I}}\left( \boldsymbol{x} \right){{\mathbb{G}}^{0\left( MM \right)}}\left( \boldsymbol{x}-\boldsymbol{y} \right):\left[ \Delta \boldsymbol{M} \left( \boldsymbol{y} \right)-{{\mathbb{D}}^{0}}:\Delta \boldsymbol{\phi} \left( \boldsymbol{y} \right) \right]d{y}d{x}}-\Delta \bar{\boldsymbol{\phi}} = 0 \\ 
\end{aligned} \right.
\end{equation}

Substituting the discretized stress and strain fields under the uniform assumption into the above equation:

\begin{equation} \label{eq:A5}
\left\{ \begin{aligned}
  & \Delta \boldsymbol{\varepsilon}^{I} + \sum\limits_{J=1}^{H}{{\boldsymbol{D}^{IJ\left( NN \right)}}\left[ \Delta \boldsymbol{N}^{J} - \mathbb{A}^{0} : \Delta \boldsymbol{\varepsilon}^{J} \right]} - \Delta \bar{\boldsymbol{\varepsilon}} = 0 \\ 
  & \Delta \boldsymbol{\phi}^{I} + \sum\limits_{J=1}^{H}{{\boldsymbol{D}^{IJ\left( MM \right)}}\left[ \Delta \boldsymbol{M}^{J} - \mathbb{D}^{0} : \Delta \boldsymbol{\phi}^{J} \right]} - \Delta \bar{\boldsymbol{\phi}} = 0 \\ 
\end{aligned} \right.
\end{equation}

where the interaction tensor $\boldsymbol{D}^{IJ\left( \sim,\sim \right)}$ is defined as:

\begin{equation} \label{eq:A6}
\left\{ \begin{aligned}
  & \boldsymbol D^{IJ\left( NN \right)} = \frac{1}{c^{I} \left| \Omega \right|} \iint_{\Omega} {\chi}^{I}(\boldsymbol x) {\chi}^{J}(\boldsymbol x) \mathbb{G}^{0\left( NN \right)}(\boldsymbol x - \boldsymbol y) \, dy \, dx \\ 
  & \boldsymbol D^{IJ\left( MM \right)} = \frac{1}{c^{I} \left| \Omega \right|} \iint_{\Omega} {\chi}^{I}(\boldsymbol x) {\chi}^{J}(\boldsymbol x) \mathbb{G}^{0\left( MM \right)}(\boldsymbol x - \boldsymbol y) \, dy \, dx \\ 
\end{aligned} \right.
\end{equation}

It can be calculated in the Fourier space as follows:

\begin{equation} \label{eq:A7}
\left\{ \begin{aligned}
  & \boldsymbol D^{IJ\left( NN \right)} = \frac{1}{c^{I} \left| \Omega \right|} \int_{\Omega} {\chi}^{I}(\boldsymbol x) \left\{ \mathcal{F}^{-1} \left[ \hat{{\chi}}^{J}(\boldsymbol \xi) : \hat{\mathbb{G}}^{0\left( NN \right)}(\boldsymbol \xi) \right] \right\} \, dx \\ 
  & \boldsymbol D^{IJ\left( MM \right)} = \frac{1}{c^{I} \left| \Omega \right|} \int_{\Omega} {\chi}^{I}(\boldsymbol x) \left\{ \mathcal{F}^{-1} \left[ \hat{{\chi}}^{J}(\boldsymbol \xi) : \hat{\mathbb{G}}^{0\left( MM \right)}(\boldsymbol \xi) \right] \right\} \, dx \\ 
\end{aligned} \right.
\end{equation}

To facilitate the self-consistent update of the reference medium parameters during the online stage, we separate the discrete frequency term and the reference medium parameter term in Green's operator as follows:

\begin{equation} \label{eq:A8}
\left\{ \begin{aligned}
  & \hat{{G}}_{khij}^{0(NN)} = \frac{1}{4 \mu_{m}^{0}} \hat{{G}}_{khij}^{1(NN)} + \frac{\lambda_{m}^{0} + \mu_{m}^{0}}{\mu_{m}^{0} \left( \lambda_{m}^{0} + 2 \mu_{m}^{0} \right)} \hat{{G}}_{khij}^{2(NN)} \\ 
  & \hat{{G}}_{khij}^{0(MM)} = \frac{1}{\lambda_{b}^{0} + 2 \mu_{b}^{0}} \hat{{G}}_{khij}^{1(MM)} \\ 
\end{aligned} \right.
\end{equation}

where,

\begin{equation} \label{eq:A9}
\left\{ \begin{aligned}
  & \hat{{G}}_{khij}^{1(NN)} = \frac{{{\delta }_{ik}}{{\xi }}_{j}{{\xi }}_{l} + {\delta }_{il}{{\xi }}_{j}{{\xi }}_{k} + {\delta }_{jl}{{\xi }}_{i}{{\xi }}_{k} + {\delta }_{jk}{{\xi }}_{i}{{\xi }}_{l}}{{\left| {\xi} \right|}^{2}}, \quad \hat{{G}}_{khij}^{2(NN)} = -\frac{{{\xi }}_{i}{{\xi }}_{j}{{\xi }}_{k}{{\xi }}_{l}}{{\left| {\xi} \right|}^{4}} \\ 
  & \hat{{G}}_{khij}^{1(MM)} = \frac{{{\xi }}_{h}{{\xi }}_{i}{{\xi }}_{j}{{\xi }}_{k}}{{\left| {\xi} \right|}^{4}} \\ 
\end{aligned} \right.
\end{equation}

The system of equations is solved using the Newton-Raphson method, where the residual of the system of equations is expressed as:

\begin{equation} \label{eq:A10}
\left\{ \begin{aligned}
  & {\mathbf{r}}^{I\left( N \right)} = \Delta {\boldsymbol{\varepsilon}}^{I} + \sum\limits_{J=1}^{K}{\boldsymbol {D}^{IJ\left( NN \right)}\left[ \Delta {\boldsymbol{N}}^{J} - {\mathbb{A}}^{0} : \Delta {\boldsymbol{\varepsilon}}^{J} \right]} - \Delta \bar{\boldsymbol{\varepsilon}} \\ 
  & {\mathbf{r}}^{I\left( M \right)} = \Delta {\boldsymbol{\phi}}^{I} + \sum\limits_{J=1}^{K}{\boldsymbol {D}^{IJ\left( MM \right)}\left[ \Delta {\boldsymbol{M}}^{J} - {\mathbb{D}}^{0} : \Delta {\boldsymbol{\phi}}^{J} \right]} - \Delta \bar{\boldsymbol{\phi}} \\ 
\end{aligned} \right.
\end{equation}

The Jacobian matrices are as follows:

\begin{equation} \label{eq:A11}
\left\{ \begin{aligned}
  & {\mathbf{J}}^{\left( N \right)} = \boldsymbol{D}^{IJ\left( NN \right)} \frac{\partial \Delta {\boldsymbol{N}}}{\partial \Delta {\boldsymbol{\varepsilon}}} + \mathrm{\boldsymbol I} - \boldsymbol{D}^{IJ\left( NN \right)} {\mathbb{A}} \\ 
  & {\mathbf{J}}^{\left( M \right)} = \boldsymbol{D}^{IJ\left( MM \right)} \frac{\partial \Delta {\boldsymbol{M}}}{\partial \Delta {\boldsymbol{\phi}}} + \mathrm{\boldsymbol I} - \boldsymbol{D}^{IJ\left( MM \right)} {\mathbb{D}} \\ 
\end{aligned} \right.
\end{equation}

\subsection{The projection-based self-consistent scheme} \label{sec:A.3}

The effective tangential stiffness of the RVE model for the partitioned H material clusters is given by:

\begin{equation} \label{eq:A12}
\left\{ \begin{aligned}
  & \bar{\mathbb{A}} = \sum_{I=1}^{H} {c^{I}} \mathbb{A}_{\text{alg}}^{I} : {\mathbb{O}}^{I} \\ 
  & \bar{\mathbb{D}} = \sum_{I=1}^{H} {c^{I}} \mathbb{D}_{\text{alg}}^{I} : {\mathbb{O}}^{I} \\ 
\end{aligned} \right.
\end{equation}

where, $\mathbb{A}_{\text{alg}}^{I}$ and $\mathbb{D}_{\text{alg}}^{I}$ represent the local constitutive laws for the incremental strain of the \textit{i}-th material cluster:

\begin{equation}  \label{eq:A13}
\left\{ \begin{aligned}
  & \mathbb{A}_{\text{alg}}^{I} = \frac{\partial \Delta \boldsymbol{N}^{I}}{\partial \Delta \boldsymbol{\varepsilon}^{I}} \\ 
  & \mathbb{D}_{\text{alg}}^{I} = \frac{\partial \Delta \boldsymbol{M}^{I}}{\partial \Delta \boldsymbol{\phi}^{I}} \\ 
\end{aligned} \right.
\end{equation}

Then, the effective tangential stiffness is projected onto a fourth-order isotropic tensor as follows:

\begin{equation} \label{eq:A14}
{{\mathbb{A}}^{0}} = \left( \mathbf{J}::\bar{\mathbb{A}} \right)\mathbf{J} + \frac{1}{5}\left( \mathbf{K}::\bar{\mathbb{A}} \right)\mathbf{K}, \quad 
{{\mathbb{D}}^{0}} = \left( \mathbf{J}::\bar{\mathbb{D}} \right)\mathbf{J} + \frac{1}{5}\left( \mathbf{K}::\bar{\mathbb{D}} \right)\mathbf{K}
\end{equation}

where,

\begin{equation} \label{eq:A15}
{{{J}}_{ijkl}} = {{\delta}_{ij}}{{\delta}_{kl}}, \quad 
{{{K}}_{ijkl}} = \frac{1}{2}\left( {{\delta}_{ik}}{{\delta}_{jl}} - {{\delta}_{il}}{{\delta}_{jk}} \right) - {{{J}}_{ijkl}}
\end{equation}

\subsection{Clustering arrangement comparison of twill woven composites} \label{sec:A.4}

For the twill woven composites in Sec. \ref{sec:3.2}, computational analyses are performed on three clustering arrangements shown in Fig.\ref{fig:F8} . Table \ref{tab:T8}) shows the errors in the homogenized properties calculated by SCA relative to the FFT results. All three arrangements show decreasing errors with increasing $H$, while maintaining relatively small error magnitudes overall. However, in terms of both mean error and errors in the homogenized properties along the principal diagonal, the A-2 has the smallest error for both the CPT and FoPT.

\begin{table}[h]
    \renewcommand*{\arraystretch}{1.5}
    \centering
    \caption{Computation error comparison between A-1, A-2 and A-3 (Unit: [\%]).}
    \label{tab:T8}      
    \resizebox{\textwidth}{!}{ 
    \begin{tabular}{l c c c c c | c c c c c}
        \hline
        & \multicolumn{5}{c|}{\textbf{CPT}} & \multicolumn{5}{c}{\textbf{FoPT}} \\
        & $H=32$ & $H=64$ & $H=128$ & $H=256$ & $H=512$ & $H=32$ & $H=64$ & $H=128$ & $H=256$ & $H=512$ \\
        \hline
        \multicolumn{11}{c}{\textbf{A-1}} \\
        $A_{11}$ & 1.666 & 1.265 & 0.829 & 0.241 & 0.122 & 1.656 & 0.797 & 0.424 & 0.113 & 0.055 \\
        $A_{12}$ & 3.776 & 2.871 & 1.423 & 0.596 & 0.294 & 3.575 & 1.275 & 0.265 & 0.303 & 0.694 \\
        $A_{33}$ & 1.323 & 0.918 & 0.666 & 0.237 & 0.124 & 1.675 & 1.138 & 0.658 & 0.365 & 0.176 \\
        $D_{11}$ & 3.039 & 2.237 & 0.873 & 0.475 & 0.241 & 3.258 & 1.655 & 1.054 & 0.494 & 0.234 \\
        $D_{12}$ & 1.764 & 0.757 & 0.288 & 0.137 & 0.074 & 2.913 & 1.525 & 1.129 & 0.606 & 0.303 \\
        $D_{33}$ & 0.104 & 0.062 & 0.048 & 0.024 & 0.008 & 0.477 & 0.091 & 0.212 & 0.270 & 0.288 \\
        Mean Err& 1.945 & 1.352 & 0.688 & 0.285 & 0.144 & 2.259 & 1.080 & 0.624 & 0.358 & 0.292 \\
        \hline
        \multicolumn{11}{c}{\textbf{A-2}} \\
        $A_{11}$ & 1.276 & 1.242 & 0.476 & 0.206 & 0.107 & 1.033 & 0.570 & 0.253 & 0.002 & 0.107 \\
        $A_{12}$ & 4.600 & 4.034 & 1.323 & 0.468 & 0.215 & 2.996 & 1.454 & 0.432 & 0.431 & 0.858 \\
        $A_{33}$ & 1.572 & 1.150 & 0.644 & 0.305 & 0.162 & 1.635 & 1.268 & 0.734 & 0.422 & 0.174 \\
        $D_{11}$ & 3.589 & 2.618 & 1.026 & 0.485 & 0.252 & 3.528 & 2.020 & 0.938 & 0.433 & 0.200 \\
        $D_{12}$ & 0.462 & 0.989 & 0.388 & 0.028 & 0.025 & 2.873 & 1.446 & 0.598 & 0.306 & 0.098 \\
        $D_{33}$ & 0.144 & 0.105 & 0.065 & 0.039 & 0.017 & 0.247 & 0.012 & 0.129 & 0.223 & 0.259 \\
        Mean Err& 1.941 & 1.690 & 0.654 & 0.255 & 0.130 & 2.052 & 1.128 & 0.514 & 0.303 & 0.283 \\
        \hline
        \multicolumn{11}{c}{\textbf{A-3}} \\
        $A_{11}$ & 1.943 & 1.242 & 0.615 & 0.358 & 0.155 & 1.473 & 0.570 & 0.225 & 0.134 & 0.081 \\
        $A_{12}$ & 5.836 & 4.034 & 1.874 & 1.047 & 0.357 & 3.427 & 1.454 & 0.418 & 0.286 & 0.668 \\
        $A_{33}$ & 1.402 & 1.150 & 0.528 & 0.254 & 0.108 & 1.701 & 1.268 & 0.650 & 0.315 & 0.106 \\
        $D_{11}$ & 4.052 & 2.618 & 1.322 & 0.779 & 0.385 & 3.755 & 2.020 & 1.308 & 1.169 & 0.458 \\
        $D_{12}$ & 0.154 & 0.989 & 0.288 & 0.100 & 0.141 & 3.229 & 1.446 & 1.126 & 1.013 & 0.294 \\
        $D_{33}$ & 0.199 & 0.105 & 0.095 & 0.065 & 0.045 & 0.393 & 0.012 & 0.122 & 0.198 & 0.231 \\
        Mean Err& 2.264 & 1.690 & 0.787 & 0.434 & 0.198 & 2.329 & 1.128 & 0.641 & 0.519 & 0.306 \\
        \hline
    \end{tabular}
    }
\end{table}

\subsection{Material properties of the practical plain woven composite} \label{sec:A.5}

All the material parameters of the practical plain woven composite are summarized in Table \ref{tab:T9}.

\begin{table}[h]
    \renewcommand*{\arraystretch}{1.5}
    \centering
    \caption{Material properties of resin and yarn phases.}
    \label{tab:T9}      
    \resizebox{\textwidth}{!}{ 
    \begin{tabular}{l p{3cm} l}
        \hline
        \textbf{Symbol} & \textbf{Value} & \textbf{Description} \\
        \hline
        $E_r$ & \raggedright 3.11 GPa & Elastic modulus of the resin \\
        $v_r$ & \raggedright 0.36 & Poisson’s ratio of the resin \\
        $\sigma^{y l}$ & \raggedright 0.0713 GPa & Yield stress of the resin \\
        $\beta_r$ & \raggedright 3.8 & Damage evolution coefficient of the resin \\
        $\rho_r$ & \raggedright 1.3 g/cm$^3$ & Density of the resin \\
        \hline
        $E_{y, 11}$ & \raggedright 128.64 GPa & Longitudinal elastic modulus of the yarn \\
        $E_{y, 22}=E_{y, 33}$ & \raggedright 9.56 GPa & Transversal elastic modulus of the yarn \\
        $v_{y, 12}=v_{y, 13}$ & \raggedright 0.28 & Poisson’s ratio of yarns on the transversal plane \\
        $v_{y, 23}$ & \raggedright 0.33 & Poisson’s ratio of yarns out of the transversal plane \\
        $G_{y, 12}=G_{y, 13}$ & \raggedright 4.47 GPa & Out-of-plane shear modulus \\
        $G_{y, 23}$ & \raggedright 3.45 GPa & In-plane shear modulus \\
        \hline
        $X^T, X^c$ & \raggedright 2710, 1630 GPa & Longitudinal tensile/compressive strength of yarns \\
        $Y^T, Y^c$ & \raggedright 0.059, 0.260 GPa & Transversal tensile/compressive strength of yarns \\
        $S^{xy}, S^{yz}$ & \raggedright 0.094, 0.080 GPa & Shear strengths of the yarn material \\
        \hline
        $\beta_\tau^f, \beta_c^f$ & \raggedright 9.0, 7.6 & Fibre damage evolution coefficients \\
        $\beta_\tau^m, \beta_c^m$ & \raggedright 4.7, 4.7 & Matrix damage evolution coefficients \\
        $\beta_s$ & \raggedright 2.8 & Shearing damage evolution coefficient \\
        $\rho_y$ & \raggedright 1.6 g/cm$^3$ & Density of the yarn \\
        \hline
    \end{tabular}
    }
\end{table}

\printbibliography

\cleardoublepage

\end{document}

\cleardoublepage

\printbibliography[title=References]

\end{document}